\documentclass[aps]{revtex4}
\usepackage[utf8]{inputenc} % If utf8 encoding
\usepackage[T1]{fontenc}    %
\usepackage[english]{babel} % English please
\usepackage{amssymb,amsmath,mathtools} % Math
\usepackage{graphicx} 
\usepackage{xcolor}
\usepackage{ulem}

\renewcommand{\Re}{\operatorname{Re}} % Real part
\renewcommand{\Im}{\operatorname{Im}} % Imaginary part

\newcommand{\MeijerG}[8][\Big]{G^{{#2 },\,{#3 }}_{{#4 },\,{#5 }} #1( \begin{smallmatrix} #6 \\ #7 \end{smallmatrix}\, #1\vert\, #8 #1)}

\newcommand{\diag}{{\rm diag\,}}
\newcommand{\sign}{{\rm sign\,}}
\newcommand{\tr}{{\rm tr\,}}

\newcommand{\U}{{\rm U\,}}

\newcommand{\RE}{{\rm Re\,}}

\newcommand{\IM}{{\rm Im\,}}
\newcommand{\eins}{\leavevmode\hbox{\small1\kern-3.8pt\normalsize1}}

\usepackage{hyperref}   % Internal hyperlinks

\numberwithin{equation}{section}

\begin{document}

\title{Universality of local spectral statistics of products of random matrices}

\author{Gernot Akemann}\email{akemann@physik.uni-bielefeld.de}\affiliation{Faculty of Physics, Bielefeld University, Postfach 100131, D-33501 Bielefeld, Germany}
\author{Zdzislaw Burda}\email{zdzislaw.burda@agh.edu.pl}\affiliation{
Faculty of Physics and Applied Computer Science,
AGH University of Science and Technology, al. Mickiewicza 30, PL-30059 Krakow, Poland}
\author{Mario Kieburg}
\email{m.kieburg@unimelb.edu.au}\affiliation{School of Mathematics and Statistics, University of Melbourne, 813 Swanston Street, Parkville, Melbourne VIC 3010, Australia}
\date{\today}

%%%%%%%%%%%%%%%%%%%%%%%%%%%%%%%%%%%%%%%%%%
\begin{abstract}  
We derive exact analytical expressions for  correlation functions of singular values of the product of $M$ Ginibre matrices of size $N$ in the double scaling limit $M,N\rightarrow \infty$.
The singular value statistics is described by a determinantal point process with a kernel that interpolates between GUE statistic and Dirac-delta (picket-fence) statistic. In the thermodynamic limit, $N\rightarrow \infty$, the interpolation parameter is given by the limiting quotient $a=N/M$.  
One of our goals is to find an explicit form of the kernel at the hard edge, in the bulk and at the soft edge for any $a$. We find that in addition to the standard scaling regimes, there is a new transitional regime which interpolates between the hard edge and the bulk. We conjecture that these results are universal, and that they apply to a broad class of products of random matrices from the Gaussian basin of attraction, including correlated matrices. We corroborate this conjecture by numerical simulations. Additionally, we show that the local spectral statistics of the considered random matrix products is identical
with the local statistics of Dyson Brownian motion with the initial condition given by equidistant positions,
with the crucial difference that this equivalence holds only locally. Finally, we have identified a mesoscopic spectral scale at the soft edge which is crucial for the unfolding of the spectrum.
\end{abstract}

%%%%%%%%%%%%%%%%%%%%%%%%%%%%%%%%%%%%%%%%%%

\maketitle

\section{Introduction}\label{intro}

Statistical properties of random matrix ensembles have been intensively studied over years and a deep understanding of the underlying principles has been achieved. Much less is known on the principles governing random matrix dynamics. There are two notable exceptions. {In the early days of random matrix theory, Dyson studied Brownian motion in matrix spaces~\cite{Dyson}; especially  the evolution of the eigenvalues of Hermitian matrices were considered starting from specific or randomised initial conditions. The evolution is given by adding Hermitian matrices made of independently identically normal distributed matrix entries to this initial matrix.} The kernel of the
{corresponding} determinantal point process, describing the evolution of the eigenvalues in the bulk, with initial condition of a fixed spectrum, was derived in~\cite{KurtBrown}. Recently, also the first step towards the formulation of a non-Hermitian version of Dyson's Brownian motion has been taken~\cite{BGNTW}.

The second example is the DMPK equation~\cite{D,MPK} for the joint-probability density function of transmission eigenvalues in a quantum wire. This equation has the form of a Fokker-Planck equation describing Brownian motion of eigenvalues propagating in a narrow wire as a function of its length which plays the role of time. The equation can be derived by constructing the transfer matrix for the whole wire as a product of independent transfer matrices for thin wire slices assuming isotropic propagation through each thin slice. The assumption of isotropy means that the flux in each ingoing channel is {in average} uniformly distributed among all outgoing channels. {One obtains} basically the same equation also for
a multiplicative stochastic model in the context of May-Wigner stability~\cite{IS}.
A common feature of Dyson's Brownian motion and of quantum transport in a quantum wire is that, mathematically, both 
are formulated as evolution equations for eigenvalues in one dimension, which 
{is} the time or the length of the system, respectively.

In the present work, we study yet another example of this type. It is an evolution of singular values of the product of random matrices. If one interprets these individual matrices as incremental transfer matrices for a time step $\Delta t$ of a system with $N$ degrees of freedom, the product of $M$ matrices can be comprehended as a transfer matrix at time $M \Delta t$. We derive an exact analytical expression for the kernels of the determinantal point processes describing the local statistics of singular values  at the hard edge, in the bulk and at the soft edge for the product of Ginibre matrices in the double scaling limit $N,M\rightarrow \infty$. 
{Previous results were obtained either at fixed $M$ when $N\to\infty$, e.g., see~\cite{kuijlaars,kieburg,kuijlaars2,LWZ,AIK,KKS,AKW,akemannstrahov}, leading to the Meijer-$G$ kernel at the hard edge \cite{kuijlaars2} or the sine- and Airy-kernel in the bulk and at the soft edge~\cite{LWZ}, respectively. Or, the limit with $N$ fixed with $M\to\infty$ was considered, e.g., see~\cite{ABK0,fk,ni,n,tutubalin,richards,reddy,kieburg}, leading to picket fence statistics~\cite{ABK0}.
A review on more recent developments is given in~\cite{ipsenakemann}.}
Here, we will consider the double scaling limit $M$ and $N\to\infty$, simultaneously.
The results in this limit  were announced in our letter~\cite{ABK} and, 
in a parallel development, partly derived in the mathematical work~\cite{LWW}.
In the present work, we will give a detailed derivation of our results in~\cite{ABK}, {where we cover the entire spectrum, including the vicinity of the hard edge and the bulk close to the soft edge, that were not contained in \cite{LWW}.
In particular, we extend previous results for standard random matrix statistics from fixed $M$ to $N\gg M$, as well as previous results for picket fence statistics at fixed $N$ to $M\gg N$, in the respective double scaling limits.}
Furthermore, we present deeper insights into these statistics, {including the issue of universality, unfolding,} and what they mean. 
{For a first work on the complex eigenvalues statistic, we refer to~\cite{LW}.}

For instance, we argue that the results hold for a broader class of multiplicative stochastic processes. This is corroborated by Monte-Carlo simulations, we have carried out, of several ensembles that include non-Gaussian ensembles as well as a certain degree of statical dependence between the matrices that are multiplied. Indeed, in a recent work~\cite{A}, it was shown that also a product of complex Jacobi matrices (truncated unitary matrices) leads to the same picture. What seems to lie behind this universality, and  came even more as surprise for us, is that the local kernels are those of the additive stochastic processes such as Dyson's Brownian motion~\cite{KurtBrown,B}. Those results of the additive processes describe the microscopic statistics of eigenvalues from an initial condition given by a non-degenerate deterministic matrix. The initial condition of the Dyson Brownian motion model is the one of the picket fence statistics (equidistant eigenvalues); for example the eigenvalue level density is
\begin{equation}\label{pf-density}
\rho_{\rm pf}(y)=\frac{1}{N}\sum_{j=1}^N\delta(y-j).
\end{equation}
Interestingly, both cases, the multiplicative as well as the additive one, yield the same limiting microscopic eigenvalue statistics in the limit $M,N\rightarrow \infty$. {This holds not only for the bulk, which has been computed for Dyson's Brownian motion in~\cite{KurtBrown}, but extends to the soft edge, too. The soft edge for the Dyson Brownian motion has not been analysed before; we will fill this gap, in the present work.} The correlations depend only on the limiting value of the parameter
\begin{equation}
a= \lim_{N\rightarrow \infty} \frac{N}{M(N)},
\label{aspect_ratio}
\end{equation}
where $a=0$ corresponds to the picket fence statistics, cf. Eq.~\eqref{pf-density}, and $a=\infty$ corresponds to the GUE local spectral statistics. The interesting and critical scaling is when the number $M$ of matrices multiplied is proportional to the matrix dimension $N$.
Hence, the additive and multiplicative processes have the same limiting local statistics and thus they belong to the same universality class, given by the interpolation of the picket fence statistics and the GUE statistics. While the bulk statistics of this {interpolating}
kernel has been derived by Johansson in~\cite{KurtBrown}, as mentioned above, the soft edge statistics has not been done, yet. We will give a brief derivation of this result and show that also at the soft edge the agreement of the kernels between the multiplicative and additive process {holds.}

Another insight we have already argued before in~\cite{ABK} and understand now very well is that actually not the ratio $N/M$ is crucial but the ratio of the average width of the distributions of individual eigenvalues about the point where one zooms in and the local mean level spacing. With the case of products of independent complex Ginibre matrices one can quantify this by the broadened picket fence spectrum, i.e., $0<a\ll1$, where the level density is not any more a sum of Dirac delta functions~\eqref{pf-density}, but a sum of log-normal distributions~\cite{ABK0}
\begin{equation}
\rho_{Y}(y) = \frac{1}{N}\sum_{j=1}^N \frac{1}{\sqrt{2\pi \sigma_j^2}y} \exp\left[ - \frac{({\rm ln}(y) - \bar{\lambda}_j)^2}{2\pi \sigma_j^2} \right] ,
\label{r1}
\end{equation}
with mean and standard deviation
\begin{equation}
\bar{\lambda}_j = \frac{\psi(j)}{2} \ , \quad \sigma_j = \sqrt{\frac{\psi'(j)}{4M}} .
\label{lsig}
\end{equation}
The Digamma function $\psi(z)=\partial_z{\rm ln}\Gamma(z)$, with $\Gamma(z)$ being the Gamma function, plays a crucial role 
for the {Lyapunov exponents of products of} Ginibre matrices. This may change for other matrix ensembles. 
{Yet, it has been recently observed~\cite{AGN}, that the asymptotic behavior of the width $\sigma_j\approx1/\sqrt{4Mj}$ [compare~\eqref{asymptotic.Digamma} below, seems to be universal, as for products of real and complex Wigner matrices the zeros of the characteristic polynomial and the positions for large Lyapunov exponents match those of the corresponding Ginibre matrices.} 

{A consequence of~\eqref{r1} is that the general width-to-spacing ratio of two consecutive eigenvalues at the mean positions $\bar{\lambda}_j $ and $\bar{\lambda}_{j+1} $ is~\cite{ABK,ABK0}
\begin{equation}
\mbox{WSR}_j = \frac{1}{2} \frac{\sigma_{j+1} + \sigma_{j}}{\bar{\lambda}_{j+1} -\bar{\lambda}_j } ,
\label{WSRdef}
\end{equation}
and equation~\eqref{aspect_ratio} has, then, to be replaced by $a=\mbox{WSR}_j ^2$ as the overlap of the distributions of the individual eigenvalues varies in the position $\bar{\lambda}_j $.} Hence, the hard edge about the origin will always exhibit picket fence statistics while for large $j\gg1$ the  transition parameter simplifies to $a\approx j/M$, which follows from the asymptotic expansions~\cite[6.3.18 and 6.4.11]{ASbook}
\begin{equation}\label{asymptotic.Digamma}
 \psi(z)= {\rm ln}(z)-\frac{1}{2z}+\mathcal{O}\left(\frac{1}{z^2}\right)\quad{\rm and}\quad
 \psi^{(l)}(z)=(-1)^{l-1}\left[\frac{(l-1)!}{z^l}+\frac{l!}{z^{l+1}}+\mathcal{O}\left(\frac{1}{z^{l+2}}\right)\right]\quad{\rm for}\quad |z|\gg1.
\end{equation}
We will see in the derivations in the ensuing sections that indeed $j/M$ is highly important in the bulk and it only happens 
at the soft-edge that $M/N$ takes the role of the
transition parameter.

As a final insight, we have found that close to the soft edge, but still in the bulk, a mesoscopic scale of spectral statistics arises. 
The microscopic statistics will be not affected and {continues to} agree with the bulk statistics, especially the {interpolating}
kernel between picket fence and GUE statistics still applies.
{However, here} 
the unfolding of the spectrum deviates from the bulk unfolding. We have already mentioned this observation in~\cite{ABK} but {at that time} did not find an analytical way to 
{derive} the proper unfolding. In the present work, we have filled in this gap. This insight is valuable and important because it gives the proper unfolding 
and allows for the identification of the universality of results. It has also  relevance for unravelling the conundrum that the macroscopic level density seems to follow always the same law, and never shows a square root behaviour at the soft edge, albeit for $N\gg M$ it is known that locally one finds the Airy-kernel, whose asymptotic form into the bulk describes a square root. This narrow region at the soft edge comprises a tail made of a certain number eigenvalue contributions. In the present work we estimate the fraction of eigenvalues which contribute 
{to be} of order $N/M$. When $M\gg N$ this tail is not present. The eigenvalues in this narrow tail have been formerly not considered and we have now derived their corresponding mesoscopic level density.

The present article is structured as follows. In Sec.~\ref{sec:prel} we briefly review the determinantal point process of the product of $M$ complex Ginibre matrices. Especially, we give two particular representations of the kernel that will be the starting point of our analysis. These representations are derived in Appendix~\ref{app:main}. Before we go over to studying the local spectral statistics in the bulk (Sec.~\ref{sec:bulk}), at the hard edge (Sec.~\ref{sec:hardedge}) and at the soft edge (Sec.~\ref{sec:softedge}), we first derive the proper unfolding for the various double scaling limits, in Sec.~\ref{sec:unfolding}. In this chapter we also unveil that there is a mesoscopic spectral regime close to the soft edge. So in addition to the discussion presented in~\cite{ABK}, we are now able to unfold the spectrum at the soft edge  
analytically. Our claim that these local spectral statistics are universal is corroborated by the Monte-Carlo simulation of several matrix ensembles including non-Gaussian as well as correlated matrices. These simulations are explained and discussed in Sec.~\ref{sec:universality}. 
{Prior to }
that we dedicate one section to the discussion of a puzzling duality between local statistics for matrix products and Dyson Brownian motion Sec. \ref{sec:duality}. {Therein, we also derive the new result of the local soft edge kernel for Dyson's Brownian motion with the picket fence spectrum as its initial condition.} In Sec.~\ref{sec:conclusio}, we summarise our findings and give an outlook on open problems.
{Further technical details are collected in Appendices \ref{app:sp} to  \ref{app:z0zs}.}

%%%%%%%%%%%%%%%%%%%%%%%%%%%%%%%%%%%%
\section{Preliminaries}\label{sec:prel}

Consider the  discrete-time evolution of an open physical system with $N$ degrees of freedom. The state of the system at time $t$ is described by an $N$-dimensional state vector $| v \rangle_t$ that evolves according to a recursive equation 
$|v\rangle_{t} = X_t | v \rangle_{t-1}$, with a transfer matrix $X_t$. The map between an initial state $|v \rangle_0$ and the state $|v \rangle_{M}=X|v \rangle_0$ after $M$ steps is given by the evolution operator
\begin{equation}
X = X_M \cdots X_1 .
\label{product}
\end{equation}
Let us assume that the transfer matrices can be modelled by random matrices. 
Since the system is open, the evolution is non-unitary so that, e.g., the norm of a state is not conserved.

We are interested in the singular value statistics of $X$ or, equivalently, in the eigenvalue statistics of the associated Hermitian operator 
\begin{equation}
Y=X^\dagger X,
\label{Y}
\end{equation}
which controls the growth of the norm $\langle v | v \rangle_M = \langle v | Y| v \rangle_0$.
The eigenvalue statistics of the product matrix $Y$ is in one-to-one correspondence with the statistics of the Lyapunov matrix 
\begin{equation}
L = \frac{1}{2M} \ln\Big[\left(X_M \cdots X_1\right)^\dagger X_M \cdots X_1\Big] =
\frac{1}{2M} \ln [Y] .
\label{LLyapunov}
\end{equation}
We concentrate on the thermodynamic limit $N\rightarrow \infty$, but at the same time assume that the number of matrices (time steps) in the product is an increasing function of the matrix size $M = M(N)$.

The microscopic spectral statistics of the Hermitian operator $Y$ is expected to be universal for a large class of transfer matrices, including the case of independent matrices $X_j$ with indepedent normal random variables as entries,   
as we shall argue later. Consequently, it is useful to consider an ensemble from this class which is analytically tractable. To be more precise, we assume that the transfer matrices, $X_t$, $t=1,\ldots, M$, are identically distributed 
independent complex Ginibre matrices~\cite{g} with i.i.d. Gaussian elements, i.e.,
\begin{equation}\label{Ginibre-P}
P(X_j)=\frac{\exp[-\tr X_jX_j^\dagger]}{\pi^{N^2}},\qquad X_j\in\mathbb{C}^{N\times N}.
\end{equation}
As shown in~\cite{AKW,AIK}, 
this ensemble 
is completely solvable, in the sense that all eigenvalue correlation functions of $Y$ of any order $k=1,\ldots, N$ can be given in closed, explicit expressions for any $M$ and $N$. 
They form a so-called determinantal point process~\cite{Kurt}, with the joint probability density of eigenvalues $y_1,\ldots,y_n>0$ of $Y$ given by 
\begin{equation}\label{DPP}
P_{Y,N}(y_1,\ldots,y_N) = \frac{1}{N!}\det \left[K_Y(y_i,y_j)\right]_{i,j=1,\ldots,N}\ .
\end{equation}
Its kernel $K_Y$ is given by \eqref{G-kernel} \cite{AKW,AIK}, and in Appendix~\ref{app:main} it is  shown to be equivalent to the forms 
\eqref{eq:main1a} and \eqref{eq:main1b} given below. 
The $k$-point correlation functions of such a point process
take an elegant determinantal form~\cite{Kurt}, as well, 
\begin{eqnarray}\label{k-point}
R_{Y,k}(y_1,\ldots,y_k) &\equiv& \frac{N!}{(N-k)!} \int_0^\infty dy_{k+1}\cdots \int_0^\infty dy_N\ P_{Y,N}(y_1,\ldots,y_N) 
\nonumber\\ 
&=&\det \left[K_Y(y_i,y_j)\right]_{i,j=1,\ldots,k}. 
\end{eqnarray}
As an example,  the normalised level density is given by $\rho_Y(y)=R_{Y,1}(y)/N=K_Y(y,y)/N$.

The first subscript in the above quantities indicates that in this case the correlation functions (kernel) are meant for the eigenvalues of the matrix $Y$. We use this convention throughout our work to distinguish between quantities for matrices $Y$, $L$ and others that will be discussed.

We would like to underline that the correlation functions \eqref{k-point} are invariant with respect to {an equivalence} 
transformation of the kernel 
\begin{equation}\label{gauge}
K_Y(x,y) \rightarrow \frac{g(x)}{g(y)} K_Y(x,y) 
\end{equation}
where $g(x)$ is a non-singular function. We shall use this invariance several times to simplify the form of the kernels.

The kernel $K_Y$ can be expressed in an explicit way 
in terms of Meijer-G functions, see~\cite{AKW,AIK} and Appendix~\ref{app:main} for details. Here, we use two equivalent and closely related representations which are particularly well suited for the various double scaling limits in $M,N \rightarrow \infty$ to be taken,
\begin{equation}
K_Y(x,y) =  \frac{1}{y}\sum_{j=0}^{N-1} \int_{\gamma_t} \frac{dt}{2\pi i} \frac{\sin (\pi (j-t))}{\pi (j-t)}\,e^{i\pi {\rm sign}[\Im(t)]t} \exp\left[-\mathcal{S}(j;x)+\mathcal{S}(t;y)\right]
\label{eq:main1a}
\end{equation}
and
\begin{equation}
K_Y(x,y) = \frac{1}{y}\int_{\gamma_t} \frac{dt}{2\pi i} \int_{\gamma_s}\frac{d s}{2\pi i} \frac{1}{s-t}\frac{\sin(\pi t)\ e^{i\pi {\rm sign}[\Im(t)]t}}{\sin(\pi s)\ e^{i\pi {\rm sign}[\Im(s)]s}}\exp\left[-\mathcal{S}(s;x)+\mathcal{S}(t;y)\right]
\label{eq:main1b}
\end{equation}
where 
\begin{equation}
\begin{split}
\mathcal{S}(z;\alpha)=&-i\pi\, {\rm sign}[\Im(z)]z
-{\rm ln}[\alpha]z+(M+1){\rm ln} [\Gamma(1+z)]+ {\rm ln}\left[\Gamma(N-z)\right],
\end{split}
\label{eq:main2}
\end{equation}
with $\alpha=x,y$ and $z=j,t,s$. The term $\mathcal{S}(z;\alpha)$ is called action in the remainder of the paper. The sign function is given by $\sign(\chi)=\chi/|\chi|$ for $\chi\in\mathbb{R}\setminus\{0\}$ and vanishes for $\chi=0$. The contour $\gamma_t$ is an integration parallel to the imaginary axis along $c+i\mathbb{R}$, with $-1<c<0$ chosen such that it does not cross the closed contour $\gamma_s$ in \eqref{eq:main1b}. The contour $\gamma_s$ encircles the closed interval $[0,N-1]$  counter-clockwise.
These formulas are derived in Appendix~\ref{app:main}, cf. 
\cite{LWW} for \eqref{eq:main1b}.

We would like to highlight that the prefactors of the exponentials in \eqref{eq:main1a} and \eqref{eq:main1b} do not grow or shrink exponentially. They  have only simple poles and zeros. Thence, they do not contribute in the saddle point equation when making an asymtotic expansion. Moreover, we would like to emphasise, that although the splitting into the exponents and prefactors is non-analytic, the integrand as a whole is a meromorphic function, see also \eqref{interm.kernel2} and \eqref{interm.kernel}, respectively.

Our goal is to analyse local (microscopic) properties of the kernel 
(\ref{eq:main1a}) in the double scaling limit $M,N\rightarrow \infty$, depending on how the limit is taken in terms of $M=M(N)$, and where in the spectrum we zoom in.
Before we discuss the local level statistics, let us 
derive in detail the relevant results on the macroscopic level density for the product of $M$ Ginibre matrices,  {in the next Section \ref{sec:unfolding}}. This preparation is necessary in order to take the local limits, where we have to unfold with respect to the macroscopic or mesoscopic level density.

%%%%%%%%%%%%%%%%%%%%%%%%%%%%%%%%%%%%%%%
\section{Macroscopic, Mesoscopic Level Density and Unfolding}\label{sec:unfolding}

In this section, we will use a saddle point analysis to determine the macroscopic level density. It is a key ingredient for the following discussion of the local statistics in the bulk and at the edges. In the discussion we will distinguish two cases, which differ in how the saddle point scales with $M$, in Subsection~\ref{sec:densityggM} and~\ref{sec:densityleqM}. We will also define what we mean by the mesoscopic density and explain when it occurs in Subsection~\ref{sec:mm}. In particular, it will be used to unfold the spectrum at the soft edge. 

For the macroscopic level density we start from~\eqref{eq:main1b}, with $x=y>0$, and perform a saddle point analysis of the action $\mathcal{S}$ in \eqref{eq:main2}. We look for the points $z_{\rm s}$ that satisfy
\begin{equation}\label{saddlepointeqbulk}
\partial_{z_{\rm s}}\mathcal{S}(z_{\rm s};y)=-{\rm ln}(y)+(M+1)\psi(1+z_{\rm s})-\psi(N-z_{\rm s})-i\pi\,\sign[\IM(z_{\rm s})]=0\ .
\end{equation}
A solution in the upper half-plane has a complex conjugate partner $z_{\rm s}^*$ in the lower half-plane. As discussed in Appendix~\ref{app:sp}, the imaginary part $\IM[z_{\rm s}]$ of the saddle point solution is of the order 
\begin{equation}\label{Imzs-order}
\IM[z_{\rm s}] = \mathcal{O}\left(\frac{\RE[z_{\rm s}]}{M}\right).
\end{equation}
The real part $\RE[z_{\rm s}]$ lies in the interval $\RE[z_{\rm s}]\in ]-1,N[$. The lower end of the interval corresponds to the hard edge and the upper one to the soft edge
of the spectrum. We see that $z_{\rm s}$ is governed by its real part. This will be used in the following two subsections in a case by case  discussion, where $\RE[z_{\rm s}]$ is taken either much larger than $M$, in Subsection \ref{sec:densityggM}, 
or much smaller than or at most of the same order as $M$, in Subsection \ref{sec:densityleqM}. Let us underline at this point that we have not assumed how $M$ and $N$ are related in the limit $M,N\rightarrow \infty$.

The hard edge is represented by $\RE[z_{\rm s}]=0$, while the soft edge by $\RE[z_{\rm s}]=N-1$. Hence, for the above estimates we require that $\RE[z_{\rm s}]\in]0,N-1[$. While for  the contour $\gamma_s$, this is not difficult to satisfy, the contour $\gamma_t$ has to be deformed accordingly. For the analysis in the bulk  we want to stay away from the edges of the spectrum. We will thus assume for the 
limiting level density that when $N\gg1$ it holds 
\begin{equation}
\label{bulk-cond}
\RE[z_{\rm s}]\gg1 \quad \mbox{and}\quad  N-1-\RE[z_{\rm s}]\gg1\ .
\end{equation}

%%%%%%%%%%%%%%%%%%%%%%%%%%%%%%%%%%%%%
\subsection{The Case: $\mathbf{\RE[z_{\rm s}]\gg M}$}\label{sec:densityggM}

In this subsection we derive 
the limiting macroscopic
level density when $\RE[z_{\rm s}]\gg M$, and identify the proper unfolding in two parts of the bulk of the spectrum. 

We shift the contour $\gamma_t$ parallel to the real axis, in particular we only set the parameter $c=\RE[z_{\rm s}]$, such that it runs through both saddle points $z_{\rm s}$ and $z_{\rm s}^*$. As above and without loss of generality, we assume that $\IM(z_{\rm s})\geq0$.
Since also the closed contour $\gamma_s$ has to run through these two saddle points, both contours have to cross each other. Originally, in the derivation of 
Eq.~\eqref{eq:main1b} in Appendix~\ref{app:main}, the contours were chosen not to cross, in order not to pick up the pole at $1/(s-t)$. To compensate this newly created residuum we have to subtract it whenever the integration path of $t$ lies inside $\gamma_s$, so that the kernel takes the form
\begin{equation}\label{kernel-split}
K_Y(y,y) =\frac{1}{y}\left[\int_{z_{\rm s}^*}^{z_{\rm s}}\frac{dt}{2\pi i}+ \int_{\RE[z_{\rm s}]-i\infty}^{\RE[z_{\rm s}]+i\infty} \frac{dt}{2\pi i} \int_{\gamma_s}\frac{d s}{2\pi i} \frac{1}{s-t}\frac{\sin(\pi t)e^{i\pi {\rm sign}[\Im(t)]t}}{\sin(\pi s)e^{i\pi {\rm sign}[\Im(s)]s}}\exp\left[-\mathcal{S}(s;y)+\mathcal{S}(t;y)\right]\right].
\end{equation}
The first integral can be readily carried out, to give $\IM[z_{\rm s}]/\pi$. This will turn out to be the dominant contribution, as we will argue in the following. 

For the second integral in \eqref{kernel-split}, we expand around the saddle points  $s=s_0+\delta s$ and $t=t_0+i\delta t$ with any combination of $s_0,t_0=z_{\rm s},z_{\rm s}^*$. This leads to a sum of four contributions to the integral. Notice that due to the form of $\gamma_t$ the perturbation $i\delta t$ always runs parallel to the imaginary axis in the same direction as the axis. In contrast, due to the form of $\gamma_s$, the perturbation $\delta s$ is real and runs anti-parallel to the real axis for $z_{\rm s}$, and parallel for $z_{\rm s}^*$, leading to a relative minus sign.
In this expansion around the saddle point the combination of actions will be replaced by 
\begin{equation}
\begin{split}\label{saddle-expansion}
-\mathcal{S}(s;y)+\mathcal{S}(t;y)\ 
{\approx}&-\mathcal{S}(s_0;y)+\mathcal{S}(t_0;y)\\
&
-\frac{(M+1)\psi'(1+s_0)+\psi'(N-s_0)}{2}\delta s^2-\frac{(M+1)\psi'(1+t_0)+\psi'(N-t_0)}{2}\delta t^2.
\end{split}
\end{equation}
When $s_0=t_0$, that is both are either $z_{\rm s}$ or $z_{\rm s}^*$, the leading contribution cancels. In the other case, the leading part of the exponent becomes
$-\mathcal{S}(z_{\rm s};y)+\mathcal{S}(z_{\rm s}^*;y)=2i \IM \mathcal{S}(z_{\rm s};y)$, or its complex conjugate.

Before we write down the integrals, let us consider the prefactors.  For $s_0=t_0=z_{\rm s}$,
we have 
\begin{equation}
\label{st-pole}
\frac{1}{s-t}=\frac{1}{\delta s-i\delta t}\ ,
\end{equation}
and
\begin{equation}
\label{sin-ratio2}
\frac{\sin[\pi(z_{\rm s}+i\delta t)]\exp[i\pi {\rm sign}[\IM(z_{\rm s}+i\delta t)](z_{\rm s}+i\delta t)]}{\sin[\pi (z_{\rm s}+\delta s)]\exp[i\pi{\rm sign}[\IM(z_{\rm s}+\delta s)] (z_{\rm s}+\delta s)]}\approx1\ .
\end{equation}
As follows from \eqref{Imzs-order}, for $\RE[z_{\rm s}]\gg M$ both real and imaginary part of $z_{\rm s}$ are large and dominate over the perturbations $\delta s$ and $i\delta t$. Equations~\eqref{st-pole} and \eqref{sin-ratio2} obviously also hold for $z_{\rm s}\to z_{\rm s}^*$. 

In the mixed case $s_0=t_0^*=z_{\rm s}$, we obtain
\begin{equation}
\frac{1}{s-t}=\frac{1}{z_{\rm s}+\delta s-z_{\rm s}^*-i\delta t}\approx \frac{1}{2i\IM[z_{\rm s}]}\ ,
\end{equation}
or its complex conjugate for $s_0^*=t_0=z_{\rm s}$. For the sine functions we have 
\begin{equation}
\sin[\pi(z_{\rm s}+i\delta t)]e^{i\pi {\rm sign}[\IM(z_{\rm s}+i\delta t)](z_{\rm s}+i\delta t)}=\frac{1}{2i}\left(e^{i\pi(z_{\rm s}+i\delta t)}-e^{-i\pi(z_{\rm s}+i\delta t)}\right) 
e^{i\pi (z_{\rm s}+i\delta t)}\approx -\frac{1}{2i}\ ,
\end{equation}
as the second term dominates, recalling that $\IM[z_{\rm s}]>0$ is large. For the same factor with  $z_{\rm s}\to z_{\rm s}^*$ we arrive at 
\begin{equation}
\sin[\pi(z_{\rm s}^*+i\delta t)]\ e^{i\pi {\rm sign}[\IM(z_{\rm s}^*+i\delta t)](z_{\rm s}^*+i\delta t)}=\frac{1}{2i}\left(e^{i\pi(z_{\rm s}^*+i\delta t)}-e^{-i\pi(z_{\rm s}^*+i\delta t)}\right) 
e^{-i\pi (z_{\rm s}^*+i\delta t)}\approx +\frac{1}{2i}\ ,
\end{equation}
as now the first term dominates. We thus  have 
\begin{equation}
\label{sin-ratio}
\frac{\sin[\pi(z_{\rm s}+i\delta t)]\exp[i\pi {\rm sign}[\IM(z_{\rm s}+i\delta t)](z_{\rm s}+i\delta t)]}{\sin[\pi (z_{\rm s}^*+\delta s)]\exp[i\pi{\rm sign}[\IM(z_{\rm s}^*+\delta s)] (z_{\rm s}^*+\delta s)]}\approx -1\ ,
\end{equation}
as well as for its complex conjugate. 
Therefore, we can now write for the kernel 
\begin{equation}\label{kernel-split.b}
\begin{split}
K_Y(y,y) \overset{M,N\gg1}{\approx}&\frac{1}{y}\biggl\{\frac{\IM[z_{\rm s}]}{\pi}-\int_{-\infty}^{\infty} \frac{d\delta t}{2\pi} \int_{-\infty}^\infty \frac{d \delta s}{\pi}\frac{1}{\delta s-i\delta t}
\IM\left[e^{-\frac{1}{2}((M+1)\psi'(1+z_{\rm s})+\psi'(N-z_{\rm s}))(\delta s^2+\delta t^2)}\right]
\\
&\quad+\int_{-\infty}^{\infty} \frac{d\delta t}{2\pi } \int_{-\infty}^\infty \frac{d \delta s}{2\pi i}\frac{e^{2i \Im[\mathcal{S}(z_{\rm s};y)]-\frac{1}{2}((M+1)\psi'(1+z_{\rm s})+\psi'(N-z_{\rm s}))\delta s^2-\frac{1}{2}((M+1)\psi'(1+z_{\rm s}^*)+\psi'(N-z_{\rm s}^*))\delta t^2}}{2i\Im[z_{\rm s}]
}
\\
&\quad-\int_{-\infty}^{\infty} \frac{d\delta t}{2\pi } \int_{-\infty}^\infty \frac{d \delta s}{2\pi i}\frac{e^{-2i \Im[\mathcal{S}(z_{\rm s};y)]-\frac{1}{2}((M+1)\psi'(1+z_{\rm s}^*)+\psi'(N-z_{\rm s}^*))\delta s^2-\frac{1}{2}((M+1)\psi'(1+z_{\rm s})+\psi'(N-z_{\rm s}))\delta t^2}}{-2i\Im[z_{\rm s}]
}
\biggl\}.
\end{split}
\end{equation}
The second integral in the first line vanishes, as one can see in polar coordinates $\delta s-i\delta t=r e^{i\varphi}$, where the integral over the angle becomes  $\int_0^{2\pi} d\varphi e^{-i\varphi}=0$. 
In the exponents in the second and third line we may expand  the Digamma function and its derivatives via Eq.~\eqref{asymptotic.Digamma}. 
Taking into account the dominance of the real part over the imaginary one~\eqref{Imzs-order}, $\RE[z_{\rm s}]\gg \IM[z_{\rm s}]=\mathcal{O}(\RE[z_{\rm s}]/M)\gg 1$, we have 
\begin{equation}
(M+1)\psi'\left(1+z_{\rm s}\right)+\psi'\left(N-z_{\rm s}\right)\approx
\frac{M}{\RE[z_{\rm s}]}+\frac{1}{N-\RE[z_{\rm s}]}\approx\frac{M}{\RE[z_{\rm s}]} ,
\end{equation}
and analogously for $z^*_{\rm s}$. Both terms on the right hand side are small and positive, due to the conditions $\RE[z_{\rm s}]\gg M$ and $N-\RE[z_{\rm s}]\gg 1$. After performing the Gaussian integrals we thus obtain for the limiting kernel
\begin{equation}\label{kernel-split.c}
\begin{split}
K_Y(y,y) \overset{M,N\gg1}{\approx}&\frac{1}{y}\biggl[\frac{\IM[z_{\rm s}]}{\pi}-\frac{\RE[z_{\rm s}]}{2\pi M\Im[z_{\rm s}]}\cos\left[2\Im[\mathcal{S}(z_{\rm s};y)]\right]\biggl].
\end{split}
\end{equation}
Because $\RE[z_{\rm s}]/(2\pi M\Im[z_{\rm s}])$ is of order one, 
the second term in \eqref{kernel-split.c} is subleading as compared to the first one which is growing with $\Im[z_{\rm s}]\gg1$. We eventually arrive at the
following asymptotic expression for the macroscopic level density
\begin{equation}\label{level.saddle}
\rho_Y(y)=\frac{1}{N}K_Y(y,y)\overset{M,N\gg1}{\approx}\frac{\IM[z_{\rm s}(y)]}{\pi N y}.
\end{equation}
Thence, the relation between the saddle point $z_{\rm s}$ and the macroscopic level density is very simple,  
{and a remarkable relation to the corresponding resolvent is pointed out in Appendic \ref{app:resolvent}.}
Let us remark that, initially, we have not imposed any relation between $N$ and $M$, apart from both $M,N\to\infty$. However,  
the condition $\RE[z_{\rm s}]\gg M$ together with the requirement to stay away from the hard edge \eqref{bulk-cond}, $N-1-\RE[z_{\rm s}]\gg1$ $\Leftrightarrow N\gg \RE[z_{\rm s}]$, implies that this case corresponds to $N\gg M$.

In order to better understand the result of the saddle point analysis of the kernel \eqref{level.saddle}, let us investigate the implications of this scaling on the relation between the argument $y \in \mathbb{R}_+$ and the saddle point $z_{\rm s}$, as it follows directly from~\eqref{saddlepointeqbulk}. Recalling our choice $\IM[z_{\rm s}]\geq0$, 
we can  exploit the asymptotic expansion~\eqref{asymptotic.Digamma} of the  Digamma function to obtain
\begin{equation}\label{zs.def}
0\approx-{\rm ln}(y)+ (M+1){\rm ln}[1+z_{\rm s}]-{\rm ln}[N-z_{\rm s}]-i\pi\quad 
\Rightarrow
\quad y \approx \frac{z_{\rm s}^{M+1}}{N-z_{\rm s}}\gg\frac{M^{M+1}}{N-M} .
\end{equation}
The last inequality follows from $\RE[z_{\rm s}]\gg M$. We also see that the limit $y\to\infty$ coincides with $\RE[z_{\rm s}]/N\to1$ for the saddle point equation~\eqref{saddlepointeqbulk}.

%%%%%%%%%%%%%%%%%%%%%%%%%%%%%%%%%%%%%%%%%%%%%%%%%%%%
\subsection{The Case: $\mathbf{\RE[z_{\rm s}]\leq \mathcal{O}(M)}$}\label{sec:densityleqM}

In this subsection, we derive the limiting macroscopic level density and find the proper unfolding in the case $\RE[z_{\rm s}]\leq \mathcal{O}(M)$, meaning $\RE[z_{\rm s}]$ is maximally of order $M$. 
Here, we have to be more careful since we need to separate the microscopic from the mesoscopic and macroscopic scale. The microscopic behaviour is encoded in the poles of the term $1/\sin(\pi s)$ in the integration variable $s$ where the integrand~\eqref{eq:main1b} gets strongly peaked. They become visible in the integral because the imaginary part of the saddle point $\IM[z_{\rm s}]=\mathcal{O}(\RE[z_{\rm s}]/M)$ is of  order one or smaller, \eqref{Imzs-order}. For this reason, we start from the series representation of the kernel \eqref{eq:main1a}, in which these poles are explicitly evaluated. 
The deformation of the $t$-integral through the saddle point does not pose a problem, as the pre-exponential factor no longer has a pole.
The index $j$ represents now the position in the spectrum and replaces the real part $\RE[z_{\rm s}]$ of the saddle point of the $s$-integration in~\eqref{eq:main1b} considered so far.

We have to be aware that in this representation the action $\mathcal{S}(j;y)$ cannot be minimised at the saddle point $z_{\rm s}$ since $j$ is an integer, $j=0,\ldots,N-1$. Therefore, we consider only the real part of the saddle point equation~\eqref{saddlepointeqbulk} which is
\begin{equation}\label{saddle-real}
{\rm ln}(y)=(M+1)\psi[1+z_{0}]-\psi[N-z_{0}]
\end{equation}
with $z_0\in]-1,N[$. Those integers $j$ that are closest to $z_0$ will contribute the most, that is why we take $z_0$ to be real. 

The uniqueness of the solution for $z_0$ is slightly simpler than for $z_{\rm s}$. For the known monoticity property of the Digamma function of real argument, cf. \cite{NIST} Section 5.3 (i) and see~\eqref{re-Digamma} for $t=0$,  it is clear that 
the right hand side of \eqref{saddle-real} is strictly increasing on $]-1,N[$ and goes to $-\infty$ for $z_0\to-1$ and to $+\infty$ for $z_0\to N$. Therefore there is a unique solution which can be given in an integral form over the Heaviside  step function,
\begin{equation}\label{z0-def}
z_0=\int_{-1}^N  \Theta\Big({\rm ln}(y)-(M+1)\psi[1+z]+\psi[N-z]\Big)dz-1.
\end{equation}
 Since Eq.~\eqref{saddle-real} gives a one-to-one relation between $y$ and $z_0$ and, hence, the summation index $j$, we have already found the proper unfolding in  $N$ and $M$, given we can show that the contributions of the index $j$ are tightly concentrated around $z_0$.

Before we come to this, let us discuss the scaling bound of $z_0$. It follows from Eq.~\eqref{saddle-real} by expanding the Digamma function, assuming that also $z_0$ is large, and exponentiating the equation 
\begin{equation}
\label{B scaling bound y}
y\approx \frac{z_0^{M+1}}{N-z_0}\leq \mathcal{O}\left(\frac{M^{M+1}}{N-M}\right)\ ,
\end{equation}
given that not only $\RE[z_{\rm s}]\leq \mathcal{O}(M)$ but also  $z_0\leq\mathcal{O}(M)$ which is consistent with \eqref{zs.def} which was the scaling bound in the opposite regime. We will make use of this insight in the ensuing discussion.

To decide which summands in~\eqref{eq:main1a} eventually contribute, we need to study the difference of {the} exponents, 
\begin{equation}\label{difference}
\begin{split}
\Delta=&\mathcal{S}(j;y)-\mathcal{S}(z_0;y)\\
=&(M+1)\left[{\rm ln}\left(\frac{\Gamma[1+j]}{\Gamma[1+z_0]}\right)-\psi[1+z_0](j-z_0)\right]+{\rm ln}\left(\frac{\Gamma[N-j]}{\Gamma[N-z_0]}\right)+\psi[N-z_0](j-z_0)\geq 0
\end{split}
\end{equation}
for $z_0\in]-1,N[$ and $j=0,1,\ldots,N-1$. Here, we have inserted already \eqref{saddle-real}.
The condition to stay in the bulk of the spectrum is 
\begin{equation}
\label{bulk-cond2}
z_0\gg1 \quad \mbox{and}\quad  N-z_0\gg1\ ,
\end{equation}
as $z_0=-1$ corresponds to the hard edge at $y=0$ and $z_0=N-1$ to the soft edge at
$y=\infty$. Note, that the maximal value of $j=N-1$ leads to a natural upper bound of the spectrum which is the soft edge. This means we also consider $j\gg1$ and $N-1-j\gg1$.

We can rewrite \eqref{difference} as
\begin{equation}\label{difference.b}
\begin{split}
\Delta
=&(\delta j)^2\int_0^1d\lambda(1-\lambda)\left[(M+1)\psi'(1+z_0+\delta j\lambda)+\psi'(N-z_0-\delta j\lambda)\right],
\end{split}
\end{equation}
where $\delta j=j-z_0$. This equation can be readily derived via integration by parts. 
{In the leading order when $|\delta j|\ll z_0$,} we get $\Delta \approx (\delta j)^2 \left( (M+1) \psi'(1+z_0) +
\psi'(N-z_0)\right)/2 \approx (\delta j)^2 M/2z_0$, for $z_0$ far from the hard and
soft edge. We have neglected higher order terms 
$(\delta j)^k$, for $k=3,4,\ldots$ Using this approximation we can find out how many summands about $z_0$ contribute, namely
\begin{equation}\label{rec:order}
\Delta \approx \frac12 (\delta j)^2\frac{M}{z_0}\leq\mathcal{O}(1)
\quad \Leftrightarrow \quad |\delta j|\leq\mathcal{O}\left(\sqrt{\frac{z_0}{M}}\right)<\mathcal{O}(z_0).
\end{equation}
Due to strict convexity of $\Delta$ we can disregard $\delta j\geq\mathcal{O}(z_0)$ as they are strongly suppressed in the exponent~\eqref{eq:main1a}.

We can insert the scaling of $\delta j = j-z_0=\mathcal{O}(\sqrt{z_0/M})$ and of $t-z_0=i\delta t=\mathcal{O}( \sqrt{z_0/M})\geq\mathcal{O}( z_0/M)=\IM[z_{\rm s}]$ {into} the action in~\eqref{eq:main1a}, and perform a Taylor expansion
\begin{equation}
\begin{split}\label{saddle-expansion.b}
-\mathcal{S}(j;y)+\mathcal{S}(t;y)\overset{M,N\gg1}{\approx}&-\frac{M}{2z_0}(\delta j^2+\delta t^2)-i\pi\sign(\delta t),
\end{split}
\end{equation}
where we have exploited \eqref{bulk-cond2}. For the kernel~\eqref{eq:main1a} we obtain
\begin{equation}
K_Y(y,y) \overset{M,N\gg1}{\approx} \frac{1}{y}\sum_{j=0}^{N-1} \int_{-\infty}^\infty \frac{d\delta t}{2\pi } \frac{\sin (\pi (j-z_0-i\delta t))}{\pi (j-z_0-i\delta t)}\exp\left[-\frac{M}{2z_0}(( j-z_0)^2+\delta t^2)\right].
\end{equation}
This integral can be evaluated by first rephrasing
\begin{equation}\label{sine-identity}
\frac{\sin (\pi (j-z_0-i\delta t))}{\pi (j-z_0-i\delta t)}=\int_{-1}^1\frac{dr}{2}\ e^{i\pi (j-z_0-i\delta t)r},
\end{equation}
then, integrating over $\delta t$ and afterwards over $r$ leading to
\begin{equation}\label{error-sum-kernel}
K_Y(y,y)\overset{M,N\gg1}{\approx} \frac{1}{2\pi y}\sum_{j=0}^{N-1}\RE\left({\rm erfi}\left[\pi\sqrt{\frac{z_0}{2M}}+i\sqrt{\frac{M}{2z_0}}(j-z_0)\right]\right).
\end{equation}
The function ${\rm erfi}(x)=-i\,{\rm erf}(ix)$ is the imaginary error function.

So far we have not separated the microscopic scale from any scale that is larger than the local mean level spacing. However, we established  the relation of $j$ with $z_0$,  and of $z_0$ with $y$. In this way, we know that when staying away from the two edges at $z_0=j=0$ and $z_0=j=N-1$ there are infinitely many eigenvalues on both sides.  Therefore, the kernel becomes discretely translation invariant under the shift $z_0\to z_0+1$; in particular, we have
\begin{equation}
K_Y(y(z_0),y(z_0))\approx K_Y(y(z_0+1),y(z_0+1))\ ,
\end{equation}
which is valid for \eqref{bulk-cond2}. Here, $y(z_0)=\exp[(M+1)\psi(1+z_0)-\psi(N-z_0)]$ is defined by the saddle point equation~\eqref{saddle-real}.  {Indeed, for $z_0\ll M$ the imaginary error functions in~\eqref{error-sum-kernel} becomes a sum of Dirac delta functions $\delta(z_0-j)$ while for $z_0\propto M$ the change of the summands in a shift $z_0\to z_0+1$ will be of order $1/M$.}

From this approximate translational invariance, we can read off that on any scale larger than the mean level spacing, which is one for the variable $z_0$, the distribution in the variable $z_0$ is uniformly distributed on $[0,N-1]$. We can also rephrase this statement and say that the limiting distribution in the original eigenvalue $y$ of the product matrix $Y$ is 
\begin{equation}\label{bulk-density}
\begin{split}
\rho_{\rm bulk}(y)=&\frac{1}{N}\frac{dz_0}{dy}=\frac{1}{Ny}\frac{dz_0}{d{\rm ln}(y)}=\frac{1}{Ny}\frac{1}{(M+1)\psi'(1+z_0)+\psi'(N-z_0)}\\
\approx&\frac{z_0}{N(M+1)y}=\frac{\int_{-1/N}^1\Theta(y-\exp[(M+1)\psi(1+N \widehat{y})-\psi(N-N\widehat{y})])d\widehat{y}-1/N}{(M+1)y}
\end{split}
\end{equation}
when $\RE[z_{\rm s}]\approx z_0\leq\mathcal{O}(M)$.
The unfolded variable is given by $\widehat{y}=z_0/N$ and therefore for finite $N$ and $M$ we have 
\begin{equation}\label{unfold.leqM}
y=\exp[(M+1)\psi(1+N \widehat{y})-\psi(N-N\widehat{y})].
\end{equation}
As a final remark, we would like to emphasise that there are various spectral scales encoded in~\eqref{unfold.leqM}. {For example, when $N\widehat{y}\propto M$, the last equation can be approximated by
\begin{equation}
y=\frac{\widehat{y}}{1-\widehat{y}}(N\widehat{y})^M \exp\left[\frac{1}{2}\frac{M}{N\widehat{y}}\right]
\label{yyy}
\end{equation}
which unfolds the spectrum in the bulk.} Depending on how $N\widehat{y}$ scales with respect to $M$, one has to go to higher orders in the asymptotic series~\eqref{asymptotic.Digamma}. 
The scaling discussed in this section corresponds to a fraction $\mathcal{O}(M/N)$ of
eigenvalues, which lie close to the hard edge. 

Let us conclude this section with a remark, that the expansion about the point $z_0$ is fully justified as it covers the contributions from the saddle point $z_{\rm s}$, as discussed in Appendix \ref{app:z0zs}.

\subsection{The Case: $\mathbf{\RE[z_{\rm s}]\gg M}$ Revisited: {Unfolding}} \label{sec:mm}

In Section \ref{sec:densityggM}, we took a double scaling limit where both $N,M\to\infty$, without specifying the rate $M=M(N)$. By assuming that $\RE[z_{\rm s}]\gg M$ and  considering a part of the spectrum away from the soft and hard edge, it emerged that $N\gg M$. It is useful to look at the regime $N\gg M$ from a slightly different perspective - by exploiting a known result for the limiting level density $\rho^{(M)}(\zeta)$ of the matrix $Y/N^M$ for $N\rightarrow \infty$ \cite{TN} {at fixed $M$}. This approach was also taken in~\cite{LWW,LWZ}. For the bulk, as we shall see, we recover the results that we have discussed in Section \ref{sec:densityggM}. However, we additionally find an unfolding map close to the soft edge that is very useful for the analysis of the soft edge statistics, that we carry out in Section \ref{sec:softedge}.

\begin{figure}[t!]
 \centering
\includegraphics[width=0.6\textwidth]{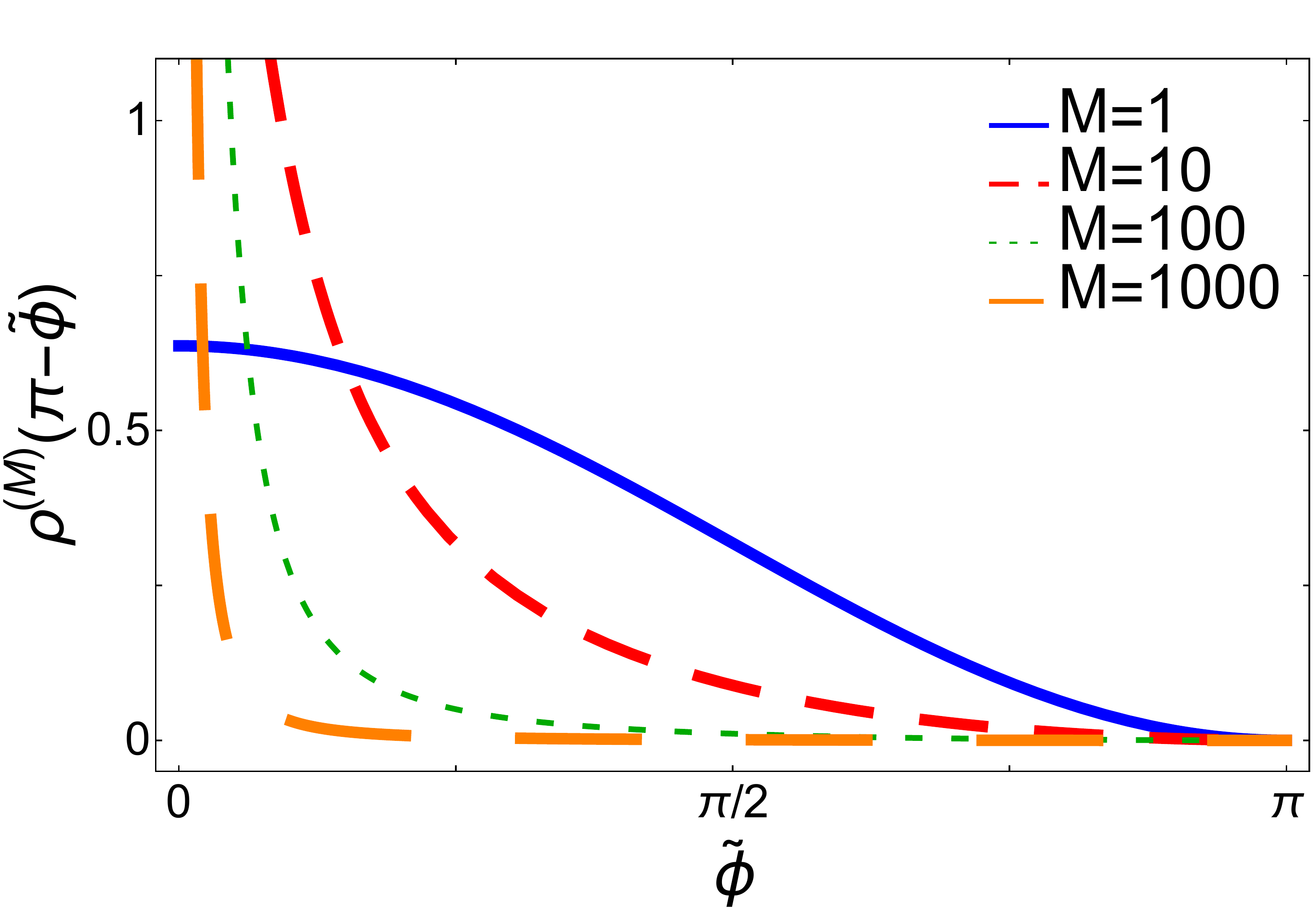}
 \caption{The macroscopic level density~\eqref{level.finiteM.phi} obtained in the limit $N\to\infty$ for various values of a fixed number of matrices $M$ multiplied. For increasing $M$ the spectrum concentrates at $\tilde{\phi}=\pi-\phi=0$ which {constitutes} 
 the deep bulk of the spectrum. Yet, the tail supported inside the interval $]0,\pi[$ still comprises roughly $N/M$ eigenvalues which is finite when $N=\mathcal{O}(M)$ and even creates a mesoscopic, soft bulk of eigenvalues when $N\gg M$. We have reflected the angle so that the hard edge lies at the origin which agrees with the picture of the hard edge of the product matrix $Y$.}
 \label{fig:rhophi}
\end{figure}

The eigenvalue density $\rho^{(M)}(\zeta)$ is given by the following parametrisation \cite{TN}
\begin{equation}\label{level.finiteM}
\rho^{(M)}(\zeta) = \frac{1}{\pi\zeta} \frac{\sin[\phi/(M+1)]\sin[\phi]}{\sin[M\phi/(M+1)]} \ ,\quad {\rm with}\quad
\zeta=\frac{\sin^{M+1}\left[\phi\right]}
{\sin[\phi/(M+1)]\sin^{M}[M\phi/(M+1)]}\quad{\rm and}\quad \phi \in ]0,\pi[.
\end{equation}
It has a support $\zeta \in [0,(M+1)^{M+1}/M^M]$.
Note that we have rescaled the parameter $\phi$ by $1/(M+1)$ compared to \cite{TN}. The hard edge located at $\zeta=0$ corresponds to $\phi=\pi$, whereas the soft edge at $\zeta=(M+1)^{M+1}/M^M$ corresponds to $\phi=0$.
It is convenient to consider two different limits of the level density~\eqref{level.finiteM}, that we call deep bulk (db) and soft bulk (sb). The corresponding unfolded variables will be specified for these two limits in the next two subsections. This can be made transparent when we give \eqref{level.finiteM} in terms of the angle $\phi$, instead of the eigenvalue $\zeta$ of the matrix $Y/N^M$. In particular we consider
\begin{equation}\label{level.finiteM.phi}
\rho^{(M)}(\phi) = \frac{1}{\pi} \frac{\sin[\phi/(M+1)]\sin[\phi]}{\sin[M\phi/(M+1)]}\left[\frac{1}{(M+1)\tan[\phi/(M+1)]}+\frac{M^2}{(M+1)\tan[M\phi/(M+1)]}-\frac{M+1}{\tan[\phi]}\right]\ {\rm with}\ \phi\in]0,\pi[,
\end{equation}
where we have multiplied the Jacobian from the transformation $\zeta\to\phi$. 
In Figure~\ref{fig:rhophi}, we reflected this density so that the hard edge is again at the origin {on the left of the plot}.

%%%%%%%%%%%%%%%%%%%%%%%%%%%%%%%%%%%%%%%%%%%%%%%%%
\subsubsection{Macroscopic Level Density and Unfolding in the Deep Bulk}\label{sec:deepbulk}

As it can be seen from Figure \ref{fig:rhophi}, the macroscopic density concentrates at $\phi\approx \pi$ for large $M$. In order to take the large $M$ limit and derive a properly scaled level density, one needs to magnify the scale close to $\pi$. We do this by introducing a new variable $\theta$ as follows $\phi=\pi-\theta/M$ with $\theta\in[0,\pi M]$. A straighforward but 
lengthy calculation using a Taylor expansion of \eqref{level.finiteM.phi} in $1/M$ leads to
\begin{equation}\label{density-db}
\frac{1}{M}\rho^{(M)}\left(\pi-\frac{\theta}{M}\right) \overset{M\gg1}{\approx}\frac{\pi}{(\theta+\pi)^2}=\rho_{\rm db}(\theta). 
\end{equation}
This scaling limit 
is called deep bulk limit (db)
because almost all eigenvalues  for $N\gg M\gg1$ are described by the density 
$\rho_{\rm db}$ \eqref{density-db}. Obviously, the part of the spectrum captured by this level density does not include the soft edge. As we shall see, it does not actually include the hard edge, either.

The corresponding unfolding in $N$ and $M$, starting  from our original eigenvalues  of the matrix $Y$ and expanding the variable $\zeta$ in \eqref{level.finiteM} to higher order in $1/M$, leads to 
\begin{equation}
y=N^M\zeta=N^M\frac{\sin^{M+1}\left[\pi-\theta/M\right]}
{\sin[(\pi-\theta/M)/(M+1)]\sin^{M}[(M\pi-\theta)/(M+1)]}\overset{M\gg1}{\approx}\frac{e\,\theta}{\pi}\left(N\frac{\theta}{\pi+\theta}\right)^M. 
\end{equation}
Thus, unfolding is given by the change of variables $\widehat{y}=1-\pi/(\theta+\pi)=\theta/(\theta+\pi)\in]0,M/(M+1)[$. In terms of $y$ it reads
 \begin{equation}\label{unfolding-bulk}
 y=e\frac{\widehat{y}}{1-\widehat{y}}\left(N\widehat{y}\right)^M\gg \frac{M^{M+1}}{N-M}\quad \Rightarrow \quad \widehat{y}\gg M/N .
 \end{equation}
 The scaling bound of $y$ in \eqref{zs.def} implies a scaling bound for the unfolded spectral variable $\widehat{y}$, given in the last inequality.
A similar scaling as in \eqref{unfolding-bulk} has been found  in~\cite{ABK}, apart from the prefactor $e\widehat{y}/(1-\widehat{y})$. {Note also that the unfolding (\ref{unfolding-bulk}) differs by an exponential factor from the unfolding (\ref{yyy}) as the scale of $\RE[z_{\rm s}]$ is different. The origin of this breakdown lies in the difference of $z_0$ and $z_{\rm s}$ which are not close any more when $\RE[z_{\rm s}]\gg M$ such that the full unfolding formula~\eqref{unfold.leqM} breaks down. However, the factor becomes irrelevant for the macroscopic spectral statistics in 
the limit $M\rightarrow \infty$.}
Such prefactors do not have any impact on the resulting unfolding of the kernel in the double limit $M,N\rightarrow \infty$, because in order to take this limit one first has to take the $M$th root and then rescale it by $1/N$, i.e., $y^{1/M}/N=[e\widehat{y}/(1-\widehat{y})]^{1/M}\widehat{y}$. So we see that the root of the prefactor converges to $1$ 
as long as we stay away from the hard edge at $\widehat{y}=0$ and from the soft edge $\widehat{y}=1$. Nevertheless, the fact that something different happens at the edges is a hint that a new mesoscopic scale may appear.

Let us quantify this by computing the normalisation of the macroscopic density \eqref{density-db}:
\begin{equation}
\int_0^{\pi M}d\theta \rho_{\rm db}(\theta)= \left.-\frac{\pi}{\theta+\pi}\right|_0^{\pi M}= 1-\frac{1}{M+1}\ .
\label{miss-soft}
\end{equation}
Thence, we are missing a fraction of the order of $1/(M+1)$ eigenvalues, which clearly vanishes only when $M\to\infty$.
The missing eigenvalues  are located close to the soft edge because $\theta$ cannot become arbitrarily large, as $\phi=\pi-\theta/M\in[0,\pi]$. This is the reason why we have to separately study a mesoscopic scaling regime close to the soft edge. We will do this in the next subsection.
 
There are also missing eigenvalues close to the hard edge at $\widehat{y}=\theta=0$. Indeed, the derivation above breaks down when the condition $\RE[z_{\rm s}]\gg M$ is violated. To see this we translate the scaling bound \eqref{unfolding-bulk} to $\theta$,
\begin{equation}
\theta\gg \frac{\pi M}{N-M}\ .
\end{equation}
Therefore, $\theta$ cannot become too small. The fraction we are missing is given naively by $M/N$, as follows from 
\begin{equation}
\int_{\pi M/(N-M)}^{\pi M}d\theta \rho_{\rm db}(\theta)= \left.-\frac{\pi}{\theta+\pi}\right|_{\pi M/(N-M)}^{\pi M}= \frac{N-M}{N}-\frac{1}{M+1}=1-\frac{M}{N}-\frac{1}{M+1} ,
\label{miss-hard}
\end{equation}
where we have corrected the integral~\eqref{miss-soft} by modifying its lower bound.
This also nicely highlights that {for $M\gg N$} the entire spectrum has to be dealt with in a different way than it is done in this subsection.

%%%%%%%%%%%%%%%%%%%%%%%%%%%%%%%%%%%%%%%%%%%%%%%%
\subsubsection{Mesoscopic Level Density and Unfolding Close to the Soft Edge}\label{sub:meso}

The aim of this subsection is to work out an unfolding map {close to}
the soft edge.
To that end we take a point-wise limit of $\rho^{(M)}(\phi)$ \eqref{level.finiteM.phi}
 with $\phi$ fixed and $M\to\infty$. This leads us to define a level density $\rho_{\rm sb}$ in the soft bulk (sb). By expanding \eqref{level.finiteM.phi} in $1/M$, we obtain
\begin{equation}\label{density-sb}
\rho^{(M)}(\phi) \overset{M\gg1}{\approx}\frac{1}{M\pi}\left[1-\frac{2\phi}{\tan[\phi]}+\frac{\phi^2}{\sin^2[\phi]}\right]+\mathcal{O}\left(\frac{1}{M^2}\right)=\rho_{\rm sb}(\phi)+\mathcal{O}\left(\frac{1}{M^2}\right).
\end{equation}
First of all, the density $\rho_{\rm sb}$ vanishes like $1/M$. Secondly, it is not integrable at the hard edge $\phi=\pi$, because it diverges there like $1/(\pi-\phi)^2$, from the last term. 
However, the hard edge is not of our interest in this section.

The corresponding parametrisation in $N$ and $M$ of the original eigenvalues $y$ of the matrix $Y$ follows from expanding $\zeta$ in \eqref{level.finiteM} in powers of 
$1/M$
\begin{equation}\label{relation.newcoord}
y=N^M\zeta
\overset{M\gg1}{\approx}
 (M+1)N^M\frac{\sin(\phi)}{\phi}\exp\left[\frac{\phi}{\tan(\phi)}\right].
\end{equation}
The corresponding level density reads
\begin{equation}\label{rho.phi.sb}
\rho_Y(y)=\rho_{\rm sb}(\phi)\left|\frac{\partial\phi}{\partial y}\right|=\frac{\phi(y)}{\pi  (M+1) y}\ ,
\end{equation}
which follows easily by first computing $\partial y/\partial \phi$ from \eqref{relation.newcoord}.

{
We can also go back to the deep bulk by letting $\phi\to\pi$. When we scale this limit  like $\phi=\pi-(\pi+\theta)/M$, due to the divergence of  $\rho_{\rm sb}(\phi)\sim \pi/[M(\pi-\phi)^2]$, we can recover the density in the deep bulk \eqref{density-db}. The shift of $\theta$ by $\pi$ is a relict hinting to the order of the limits. Since we can go back to the deep bulk regime in this way, we can exclude that we have missed any other intermediate mesoscopic scaling regime.}

The unfolding in the mesoscopic regime is given by
\begin{equation}\label{unfold.sb}
\widehat{y}=\int_0^{\phi(y)}d\phi'\rho_{\rm sb}(\phi')=\frac{1}{(M+1)\pi}\left(\phi(y)-\frac{\phi(y)^2}{\tan(\phi(y))}\right).
\end{equation}
For the dependence on the original eigenvalue $y$ we need to invert the relation~\eqref{relation.newcoord} for $\phi = \phi(y)$. Formally one can write
the solution as
\begin{equation}\label{phi.int}
\phi(y)=\int_0^\pi d\phi\ \Theta\left( (M+1)N^M\frac{\sin(\phi)}{\phi}\exp\left[\frac{\phi}{\tan(\phi)}\right]-y\right)
\end{equation}
by noticing that the function on the right hand side of~\eqref{relation.newcoord} is strictly decreasing on the interval $[0,\pi]$. We shall use this unfolding in Section \ref{sec:softedge} and Section \ref{sec:universality} while discussing the soft edge local statistics. 

Finally, when combining \eqref{relation.newcoord} and \eqref{rho.phi.sb} one finds {that} the density has a square root behaviour 
at the upper edge $y_+ = N^M(M+1)^{M+1}/M^M \approx e (M+1)N^M$ of the support
\begin{equation}\label{density-near-soft}
\rho_Y(y)\overset{y_+-y\ll y_+}{=}\frac{\sqrt{2}}{\pi (M+1) y_+}\sqrt{1-
\frac{y}{y_+}}.
\end{equation}
{This square root behaviour describes a substantial portion of the spectrum when $M\leq\mathcal{O}(N)$, namely about $N/M$ eigenvalues. For instance for $N\gg M$, this number is infinitely large, indicating the onset of bulk statistics but with respect to the unfolding pointed out in this subsection which is evidently different from the one in Subsection~\ref{sec:deepbulk}.}
%%%%%%%%%%%%%%%%%%%%%%%%%%%%%%%%%

\section{Bulk Statistics}\label{sec:bulk}

The local statistics are always defined by choosing a base point $y_0\in\mathbb{R}_+$ and zooming into the vicinity of this point. Zooming in means here unfolding and this implies that the macroscopic level density becomes flat. Therefore, we first compute the saddle point $z_{\rm s}$ (with $\IM[z_{\rm s}]\geq0$) with respect to the base point $y_0$ with the aid of equation~\eqref{saddlepointeqbulk}.

%%%%%%%%%%%%%%%%%%%%%%%%%%%%%%%%%%%%
\subsection{Sine-Kernel}\label{sec:sine}

As we have seen, the corresponding level density is for $\RE[z_{\rm s}]\gg M$ given by~\eqref{level.saddle}. This already tells us what the correct unfolded variables are for the two spectral variables in the kernel~\eqref{eq:main1b},
\begin{equation}\label{unfolding.ggM}
x=y_0\exp\left[\frac{\pi}{\Im[z_{\rm s}]}\widehat x\right]\ {\rm and}\ y=y_0\exp\left[\frac{\pi}{\Im[z_{\rm s}]}\widehat y\right]\ {\rm with}\ \widehat x,\widehat y=\mathcal{O}(1)
\end{equation}
 because their corresponding density is flat, i.e.,
\begin{equation}
N\rho(y)dy=\frac{{\IM[z_{\rm s}}]}{\pi  y_0\exp\left[\pi\widehat y/\Im[z_{\rm s}]\right]} d\left(y_0\exp\left[\frac{\pi}{\Im[z_{\rm s}]}\widehat y\right]\right)
{\approx}\ d\widehat y.
\end{equation}
The dependence of $z_{\rm s}(y)$ on $\widehat{y}$ is vanishing {due to $\Im[z_{\rm s}]=\mathcal{O}(\RE[z_{\rm s}]/M)\gg1$.} Thus, we can replace $z_{\rm s}(y)$ by $z_{\rm s}(y_0)$ which is the reason why we drop its argument.

The microscpic bulk statistics  can be readily obtained by plugging the change of variables~\eqref{unfolding.ggM} into the kernel~\eqref{eq:main1b},
\begin{equation}
\begin{split}
K_Y(x,y)\frac{dy}{d\widehat{y}} =&\frac{\pi}{\Im[z_{\rm s}]}\int_{\gamma_t} \frac{dt}{2\pi i} \int_{\gamma_s}\frac{d s}{2\pi i} \frac{1}{s-t}\frac{\sin(\pi t)e^{i\pi {\rm sign}[\Im(t)]t}}{\sin(\pi s)e^{i\pi {\rm sign}[\Im(s)]s}}\exp\left[\frac{\pi\widehat x}{\Im[z_{\rm s}]}s-\frac{\pi\widehat y}{\Im[z_{\rm s}]}t\right]\exp\left[-\mathcal{S}(s;y_0)+\mathcal{S}(t;y_0)\right].
\end{split}
\label{eq:main1.ggM}
\end{equation}
The factor $dy/d\widehat y$ is the resulting Jacobian of the change of variable.
As for the level density we shift the contour $\gamma_t$ to $\RE[z_{\rm s}]+i\mathbb{R}$ while the contour $\gamma_s$ runs through $z_{\rm s}$ and $z_{\rm s}^*$. Thence, we obtain a residuum which leads to
\begin{equation}
\begin{split}
K_Y(x,y)\frac{dy}{d\widehat{y}} =&\frac{\pi}{\Im[z_{\rm s}]}\int_{z_{\rm s}^*}^{z_{\rm s}}\frac{dt}{2\pi i} \exp\left[\frac{\pi(\widehat x-\widehat y)}{\Im[z_{\rm s}]}t\right]\\
&+\frac{\pi}{\Im[z_{\rm s}]} \int_{\RE[z_{\rm s}]-i\infty}^{\RE[z_{\rm s}]+i\infty} \frac{dt}{2\pi i} \int_{\gamma_s}\frac{d s}{2\pi i} \frac{1}{s-t}\frac{\sin(\pi t)e^{i\pi {\rm sign}[\Im(t)]t}}{\sin(\pi s)e^{i\pi {\rm sign}[\Im(s)]s}}\exp\left[\frac{\pi\widehat x}{\Im[z_{\rm s}]}s-\frac{\pi\widehat y}{\Im[z_{\rm s}]}t\right]e^{-\mathcal{S}(s;y_0)+\mathcal{S}(t;y_0)}.
\end{split}
\end{equation}
The first term is the one we are looking for since it evaluates to
\begin{equation}
\begin{split}
\frac{\pi}{\Im[z_{\rm s}]}\int_{z_{\rm s}^*}^{z_{\rm s}}\frac{dt}{2\pi i} \exp\left[\frac{\pi(\widehat x-\widehat y)}{\Im[z_{\rm s}]}t\right]=\exp\left[\pi\frac{\RE[z_{\rm s}]}{\IM[z_{\rm s}]}(\widehat x-\widehat y)\right]\frac{\sin[\pi(\widehat x-\widehat y)]}{\pi(\widehat x-\widehat y)}.
\end{split}
\end{equation}
We exploit the invariance~\eqref{gauge} to get rid of the exponential factor in the expression on the right hand side by multiplying it by
$\exp\left[-\pi\RE[z_{\rm s}](\widehat x-\widehat y)/\IM[z_{\rm s}]\right]$.

The second term multiplied by this new factor vanishes then in the large $N,M$ limit. This can be seen by expanding the integration variables 
$s=s_0+\delta s$ and $t=t_0+i\delta t$
about the two saddle points $s_0,t_0=z_{\rm s},z_{\rm s}^*$. 
In particular, we employ~\eqref{saddle-expansion} for the spectral variable $y_0$. The integrals are indeed bounded because for {$s_0=t_0=z_{\rm s}$ (and similarly for  $s_0=t_0=z_{\rm s}^*$, with $i\to-i$), and we have }
\begin{equation}
\exp\left[\frac{\pi\widehat x}{\Im[z_{\rm s}]}s-\frac{\pi\widehat y}{\IM[z_{\rm s}]}t-\pi\frac{\RE[z_{\rm s}]}{\IM[z_{\rm s}]}(\widehat x-\widehat y)\right]=\exp\left[i(\widehat x-\widehat y)+\frac{\pi\widehat x}{\Im[z_{\rm s}]}\delta s-i\frac{\pi\widehat y}{\Im[z_{\rm s}]}\delta t\right]\overset{M,N\to\infty}{\longrightarrow}e^{i(\widehat x-\widehat y)}.
\end{equation}
Note that $\delta s,\delta t=\mathcal{O}(\sqrt{\IM[z_{\rm s}]})$ because of the scaling of the quadratic terms~\eqref{saddle-expansion}. For $s_0=t_0^*=z_{\rm s}$ {(and similarly for $s_0=t_0^*=z_{\rm s}^*$, with $i\to-i$)}, we have
\begin{equation}
\exp\left[\frac{\pi\widehat x}{\Im[z_{\rm s}]}s-\frac{\pi\widehat y}{\Im[z_{\rm s}]}t-\pi\frac{\RE[z_{\rm s}]}{\IM[z_{\rm s}]}(\widehat x-\widehat y)\right]=\exp\left[i(\widehat x+\widehat y)+\frac{\pi\widehat x}{\Im[z_{\rm s}]}\delta s{-i\frac{\pi\widehat y}{\Im[z_{\rm s}]}\delta t}\right]\overset{M,N\to\infty}{\longrightarrow}e^{i(\widehat x+\widehat y)}.
\end{equation}
Since these double contour integrals are multiplied by $1/\IM[z_{\rm s }]=\mathcal{O}(M/\RE[z_{\rm s}])=o(1)$ the terms vanish so that we find the celebrated sine-kernel~\cite{GUE}
\begin{equation}\label{sine-kernel}
K_{\rm sin}(\widehat{x},\widehat{y})=\lim_{M,N\to\infty} \exp\left[-\pi\frac{\RE[z_{\rm s}]}{\IM[z_{\rm s}]}(\widehat x-\widehat y)\right]\ K_Y(x,y)\frac{dy}{d\widehat{y}}=\frac{\sin[\pi(\widehat x-\widehat y)]}{\pi(\widehat x-\widehat y)}
\end{equation}
for all cases when the base point $y_0$ satisfies the scaling bounds $y_0\gg M^{M+1}/(N-M)$ and $1-y_0/(e (M+1)N^M)\gg (M/N)^{2/3}$. 
{This extends the result of \cite{LWZ} obtained when $M$ is fixed. }
The first bound shows the transition to the other double scaling limits. The latter bound at the soft edge follows from the fact that it always vanishes like a square root, particularly the mesoscopic  level density has at the soft edge  the form~\eqref{density-near-soft}.

Satisfying both bounds implies the double scaling $N\gg M\gg1$. Therefore the sine-kernel cannot always be found.

%%%%%%%%%%%%%%%%%%%%%%%%%%%%%%%%%%%
\subsection{Picket Fence Statistics in the Bulk}\label{sec:picketfence.bulk}

{For this regime, we  require the stricter condition $\Re[z_{\rm s}(y_0)]=o(M)$, rather 
than $\Re[z_{\rm s}(y_0)]\leq \mathcal{O}(M)$ which was the scaling in Subsection \ref{sec:densityleqM}.}
We choose an unfolding of the form
\begin{equation}\label{unfolding.llM}
x=y_0\exp\left[(M+1)\psi'(1+z_0)\widehat x\right]\ {\rm and}\ y=y_0\exp\left[(M+1)\psi'(1+z_0)\widehat y\right]\ {\rm with}\ \widehat x,\widehat y=\mathcal{O}(1)
\end{equation}
with $z_0$ given by~\eqref{saddle-real} with $y\to y_0$. This change indeed flattens the bulk density~\eqref{bulk-density} in the present scaling limit.

As we have seen in Appendix \ref{app:z0zs}, the saddle point $z_{\rm s}$ is infinitesimally close to the real point $z_0$. The contributing summands are those with an index $j$ that satisfies the scaling $|j-z_0|=\mathcal{O}(\sqrt{\Re[z_{\rm s}]/M})$. Because of $\Re[z_{\rm s}]\approx z_0 \ll M$, the kernel does not vanish only when $z_0=0,1,\ldots,N-1$. Indeed the Gaussian approximation~\eqref{saddle-expansion.b} of the action for the sum representation of the kernel~\eqref{eq:main1a} simplifies to a sum of Dirac delta functions. One can show this by the following computation. The kernel in the new variables is
\begin{equation}
\begin{split}
K_Y(x,y) \frac{dy}{d\widehat{y}}=& (M+1)\psi'(1+z_0)\sum_{j=0}^{N-1} \int_{\gamma_t} \frac{dt}{2\pi i} \frac{\sin (\pi (j-t))}{\pi (j-t)}e^{i\pi {\rm sign}[\Im(t)]t} \exp\left[(M+1)\psi'(1+z_0)(\widehat x j-\widehat y t)\right]\\
&\times\exp\left[-\mathcal{S}(j;y_0)+\mathcal{S}(t;y_0)\right]\\
\approx&(M+1)\psi'(1+z_0)\sum_{j=0}^{N-1} \int_{-\infty}^\infty \frac{d\delta t}{2\pi } \frac{\sin (\pi (j-z_0-i\delta t))}{\pi (j-z_0-i\delta t)}\exp\left[M\psi'(1+z_0)\big(\widehat x j-\widehat y (z_0+i\delta t)\big)\right]\\
&\times\exp\left[-\frac{M\psi'(1+z_0)}{2}((j-z_0)^2+\delta t^2)\right].
\end{split}
\end{equation}
First, we shift the integration variable $\delta t\to\delta t-i\widehat{y}$
\begin{equation}\label{kernel-pf-inter}
\begin{split}
K_Y(x,y) \frac{dy}{d\widehat{y}}\approx& (M+1)\psi'(1+z_0)\exp\left[\frac{M\psi'(1+z_0)}{2}([\widehat{x}+z_0]^2-[\widehat{y}+z_0]^2)\right]\\
&\times\sum_{j=0}^{N-1} \int_{-\infty}^\infty \frac{d\delta t}{2\pi } \frac{\sin (\pi (j-z_0-\widehat{y}-i\delta t))}{\pi (j-z_0-\widehat{y}-i\delta t)}\exp\left[-\frac{M\psi'(1+z_0)}{2}([j-z_0-\widehat{x}]^2+\delta t^2)\right].
\end{split}
\end{equation}
In the next step, we can replace the two Gaussians by two Dirac delta functions where we can evaluate the integral over $\delta t$. This leads to the result
\begin{equation}\label{pf-kernel-interp}
K_Y(x,y) \frac{dy}{d\widehat{y}}\approx \exp\left[\frac{M\psi'(1+z_0)}{2}([\widehat{x}+z_0]^2-[\widehat{y}+z_0]^2)\right]\sum_{j=0}^{N-1}\frac{\sin (\pi (j-z_0-\widehat{y}))}{\pi (j-z_0-\widehat{y})}\delta( j-z_0-\widehat{x}).
\end{equation}
The factor in front of the sum can be skipped on virtue of the invariance~\eqref{gauge} {of the kernel}. Additionally, we consider the bulk which means that $z_0-1,N-1-z_0\gg1$ so that we can extend the sum into both directions {to $\pm\infty$} in the limit $N,M\to\infty$ when splitting $j-z_0=\widehat{j}+\nu$ with $\widehat{j}\in\mathbb{Z}$ and $\nu\in]-0.5,0.5]$. The parameter $|\nu|$ is thus the distance of $z_0$ to its closest integer. Thence, we eventually arrive at
\begin{equation}\label{pf-bulk}
\begin{split}
K_{\rm pf}(\widehat{x},\widehat{y})=&\lim_{M,N\to\infty} \exp\left[-\frac{M\psi'(1+z_0)}{2}([\widehat{x}+z_0]^2-[\widehat{y}+z_0]^2)\right]K_Y(x,y) \frac{dy}{d\widehat{y}}\\
=&\sum_{\widehat{j}=-\infty}^{\infty}\frac{\sin (\pi (\widehat{j}+\nu-\widehat{y}))}{\pi (\widehat{j}+\nu-\widehat{y})}\delta(\widehat{j}+\nu-\widehat{x}).
\end{split}
\end{equation}
This is the kernel of an equidistant spectrum which looks like a picket fence {in both directions}, thus, the name picket fence statistics {(pf)}. The sine function in the summand is essential since it guarantees that no two eigenvalues lie at the same position. For instance, for the microscopic one-point and two-point functions we get
\begin{equation}
R_1(\widehat{y})=\sum_{\widehat{j}=-\infty}^{\infty}\frac{\sin (\pi (\widehat{j}+\nu-\widehat{y}))}{\pi (\widehat{j}+\nu-\widehat{y})}\delta(\widehat{j}+\nu-\widehat{y})=\sum_{\widehat{j}=-\infty}^{\infty}\delta(\widehat{j}+\nu-\widehat{y})\ ,
\end{equation}
and
\begin{equation}
\begin{split}
R_2(\widehat{x},\widehat{y})=&R_1(\widehat{x})R_1(\widehat{y})-\sum_{\widehat{j},\widehat{l}=-\infty}^{\infty}\frac{\sin (\pi (\widehat{j}+\nu-\widehat{y}))}{\pi (\widehat{j}+\nu-\widehat{y})}\frac{\sin (\pi (\widehat{l}+\nu-\widehat{x}))}{\pi (\widehat{l}+\nu-\widehat{x})}\delta(\widehat{j}+\nu-\widehat{x})\delta(\widehat{l}+\nu-\widehat{y})\\
=&R_1(\widehat{x})R_1(\widehat{y})-\sum_{\widehat{j},\widehat{l}=-\infty}^{\infty}\delta_{\widehat{j}\widehat{l}}\,\delta(\widehat{j}+\nu-\widehat{x})\delta(\widehat{l}+\nu-\widehat{y})\\
=&\sum_{\widehat{j}\neq \widehat{l}}\delta(\widehat{j}+\nu-\widehat{x})\delta(\widehat{l}+\nu-\widehat{y}),
\end{split}
\end{equation}
respectively.

Let us underline that the kernel~\eqref{pf-bulk} always holds when the real part of the saddle point solution satisfies $1\ll\RE[z_{\rm s}]=z_0\ll M$ and $N-\RE[z_{\rm s}]\gg1$. This translates into the original eigenvalue $y_0$ of the product matrix $Y$ into the form ${{\rm ln}(N y_0)\gg M}$ and $(M+1)\psi[N]-\psi[1]-{\rm ln}(y_0)\gg \max\{M/N,1\}$ meaning that one stays away from the hard and soft edge, respectively, as well as $y_0\ll (M+1)^{M+1}/[N-M]$ which represents the regime where the distributions of the individual eigenvalues do not overlap anymore. The condition $(M+1)\psi[N]-\psi[1]-{\rm ln}(y_0)\gg \max\{M/N,1\}$ follows from Eq.~\eqref{saddle-real} at the maximal value $z_0=N-1$ and taking the difference of this equation for $z_0-\delta j$ with $\delta j$ of order one. Namely, for $\delta j=\mathcal{O}(1)$ one still sees the upper boundary of the spectrum.

From these scaling boundaries we see that there is always a part of the spectrum of $Y$ whose spectral statistics follows the kernel~\eqref{pf-bulk} regardless of the relation between $M$ and $N$ in the double scaling limit. For $M\gg N$ 
{the entire bulk} is governed by {the picket fence} local spectral statistics. 
{This extends the results of \cite{ABK0} where this was shown for fixed $N$ only.}

\subsection{Critical Regime in the Bulk}\label{sec:trans.bulk}

The two bulk statistics above have a transition regime which is given by the fact that $\RE[z_{\rm s}]\approx z_0=\mathcal{O}(M)$, {which is included in the scaling $\RE[z_{\rm s}]\leq\mathcal{O}(M)$ in Subsection \ref{sec:densityleqM}.} In this regime, we unfold the variables like in~\eqref{unfolding.llM} since the level density is the same, see~\eqref{bulk-density}. Replacing $\psi'(1+z_0)$ by its asymptotic behaviour $1/z_0 + 1/(2z_0^2) + \ldots$, \eqref{asymptotic.Digamma}, we have
\begin{equation}\label{unfolding.eqM}
x=y_0\exp\left[\frac{M+1}{z_0}\widehat x\right]\ {\rm and}\ y=y_0\exp\left[\frac{M+1}{z_0}\widehat y\right]\ {\rm with}\ \widehat x,\widehat y=\mathcal{O}(1).
\end{equation}
The term $M/z_0$ is of order one for $z_0=\mathcal{O}(M)$. All other terms can be neglected. The next-to-leading one is of order $M/z_0^2 \sim \mathcal{O}(1/M)$ so it vanishes in the limit $M\rightarrow \infty$.

We can already start from the intermediate result~\eqref{kernel-pf-inter},
{which was derived for $\RE[z_{\rm s}]={o}(M)$ but still holds here as it only takes into account the computations  in Subsection \ref{sec:densityleqM} which hold for $\RE[z_{\rm s}]\leq\mathcal{O}(M)$. It} reads now
\begin{equation}
\begin{split}
K_Y(x,y) \frac{dy}{d\widehat{y}}\approx&\frac{M+1}{z_0}\exp\left[\frac{M}{2z_0}([\widehat{x}+z_0]^2-[\widehat{y}+z_0]^2)\right]\\
&\times\sum_{j=0}^{N-1} \int_{-\infty}^\infty \frac{d\delta t}{2\pi } \frac{\sin (\pi (j-z_0-\widehat{y}-i\delta t))}{\pi (j-z_0-\widehat{y}-i\delta t)}\exp\left[-\frac{M}{2z_0}([j-z_0-\widehat{x}]^2+\delta t^2)\right].
\end{split}
\end{equation}
This time we cannot replace the two Gaussians by Dirac delta functions since their variance is of order one. To carry out the $\delta t$ integral, nevertheless, we exploit the same trick as in~\eqref{sine-identity} and arrive at the result
\begin{equation}\label{kernel-trans-bulk}
\begin{split}
K_{\rm cb}(\widehat{x},\widehat{y})=&\lim_{\substack{M,N\to\infty\\z_0/M\to a}}\exp\left[-\frac{M}{z_0}(\nu+z_0)(\widehat{x}-\widehat{y})\right]K_Y(x,y) \frac{dy}{d\widehat{y}}\\
=&\frac{1}{2\pi a}\sum_{\widehat{j}=-\infty}^{\infty}\RE\left({\rm erfi}\left[\pi\sqrt{\frac{a}{2}}+i\sqrt{\frac{1}{2a}}(\widehat{j}+\nu-\widehat{y})\right]\right)\exp\left[\frac{1}{a}(\widehat{x}-\widehat{y})\widehat{j}\right].
\end{split}
\end{equation}
The subscript ${\rm cb}$ stands for critical bulk. We used the invariance \eqref{gauge}
to introduce the prefactor in the last expression.
As in the picket fence case the index $j$ is closely bound to $z_0$, $j-z_0=\widehat{j}+\nu$ must be of order one. This in combination with the fact that $1+z_0,N-1-z_0\gg1$ has allowed us to extend the sum to $\mathbb{Z}$.

The result~\eqref{kernel-trans-bulk} holds for those $y_0$ which are of the order $(M+1)^{M+1}/[N-M]$ which is the scale where the single eigenvalues start to feel their neighbouring eigenvalues. Additionally, we need to stay away from the soft edge so that it is also $(M+1)\psi[N]-\psi[1]-{\rm ln}(y_0)\gg \max\{M/N,1\}$. This condition is however of relevance only when $N=\mathcal{O}(M)$. Only then this critical regime reaches the soft edge.

The result~\eqref{kernel-trans-bulk} was already presented in~\cite{ABK}. {In Sec. \ref{sec:duality}, we give a detailed derivation showing that the same kernel is obtained
from Dyson’s Brownian motion.} There is also an alternative version of this kernel shown in~\cite{LWW} which is expressed in terms of the Jacobi theta function
\begin{equation}\label{Jacobi}
\vartheta(z;\tau)=\sum_{j=-\infty}^\infty\exp[ \pi i j^2 \tau+2\pi i j z].
\end{equation}
For this purpose, we do not carry out the $r$-integral resulting from the trick~\eqref{sine-identity} so that the kernel has the form
\begin{equation}\label{kernel-trans-bulk.ab}
\begin{split}
K_{\rm cb}(\widehat{x},\widehat{y})=&\sqrt{\frac{1}{8\pi a}}\sum_{\widehat{j}=-\infty}^\infty \int_{-1}^1 dr\ \exp\left[-\frac{1}{2a}\widehat{j}^2+i\left(\pi r-i\frac{1}{a}(-\nu+\widehat x)\right)\widehat{j}+\frac{a}{2 }\left(\pi r-i\frac{1}{a}(-\nu+\widehat{y})\right)^2\right].
\end{split}
\end{equation}
In this expression one can identify the series with the Jacobi theta function~\eqref{Jacobi} which yields~\cite{LWW}
\begin{equation}\label{kernel-trans-bulk.b}
\begin{split}
K_{\rm cb}(\widehat{x},\widehat{y})=&\sqrt{\frac{1}{8\pi a}} \int_{-1}^1 dr\ \vartheta\left(\frac{r}{2}-i\frac{1}{2\pi a}(-\nu+\widehat x);i\frac{1}{2\pi a}\right)\exp\left[\frac{a}{2}\left(\pi r-i\frac{1}{a}(-\nu+\widehat{y})\right)^2\right].
\end{split}
\end{equation}

Let us finally mention that the kernel~\eqref{kernel-trans-bulk} exhibits again a discrete translation symmetry $(\widehat{x},\widehat{y})\to(\widehat{x}+1,\widehat{y}+1)$ as does the picket fence statistics. It reflects the fact that we have indeed properly unfolded the spectrum since the averaged mean level distance is one. When taking the ratio $a=z_0/M\to0$ we indeed regain the picket fence kernel~\eqref{pf-bulk}. This can be easiest seen in~\eqref{kernel-trans-bulk.ab} where the Gaussain in $\widehat{j}$ can be replaced by a Dirac delta function and the remaining integral in $r$ is a simple exponential function leading to the sinus cardinalis in~\eqref{pf-bulk}.

The sine-kernel result~\eqref{sine-kernel} can be found when sending $a=z_0/M\to\infty$. Then, the sum~\eqref{kernel-trans-bulk.ab} in $\widehat{j}$ becomes quasi-continuous and, hence, a simple Gaussian. Carrying out this Gaussian integral in $\widehat{j}$ leads again to an exponential integrand in $r$ giving the sinus cardinalis in~\eqref{sine-kernel}.

\section{Hard Edge}\label{sec:hardedge}

For the hard edge, the Diagamma function $\psi[1+z_0]$ as well as its derivatives cannot be approximated by the asymptotic~\eqref{asymptotic.Digamma} while the Digamma function $\psi[N-z_0]$ is essentially ${\rm ln}(N)$ and its derivatives {are subleading.} 
Therefore, the  reduced saddle point equation~\eqref{saddle-real} simplifies to
\begin{equation}
\psi[1+z_0]=\frac{{\rm ln}(Ny)}{M+1}.
\end{equation}
Hence, the spectral variable scales like $y=\exp[(M+1)\psi(1+z_0)]/N$. When comparing this result with those in~\cite{ABK} one can easily notice a difference by a factor $\exp[\psi(1+z_0)]$. As long as ${\rm ln}(z_0)\ll M$, this term always vanishes in the unfolding because its $M$-th root goes to one. Yet, there will be eventually corrections when ${\rm ln}(z_0)\geq\mathcal{O}(M)$. This latter regime is not the case for the hard edge scaling. As we will see, the variable $z_0$ has to be still very close to an integer, so that the smooth part of the level density for $y$ follows 
\begin{equation}
\rho_{\rm hard}(y)=\frac{1}{N(M+1)y\ \psi'[1+\int_0^\infty ds\ \Theta(y-\exp[(M+1)\psi(1+s)]/N )]}.
\end{equation}
In principal, one can also consider this density as the mesoscopic part of the density~\eqref{bulk-density} close to the hard edge. Therefore, the unfolding is given by
\begin{equation}\label{unfolding.hard}
x=\exp\left[(M+1)\psi[1+\widehat x]\right]/N\ {\rm and}\ y=\exp\left[(M+1)\psi[1+\widehat y]\right]/N\ {\rm with}\ \widehat x,\widehat y=\mathcal{O}(1).
\end{equation}

Essentially, we can make use of the analysis of subsection~\ref{sec:densityleqM} and~\ref{sec:picketfence.bulk} since the present discussion does not differ much from it. The only but essential difference is that $\widehat{x}$ and $\widehat{y}$ are already the solutions of~\eqref{saddle-real} so that we expand the summation index $j$ about $\widehat{x}$ and the integration variable $t$ about $\widehat{y}$. Here, we begin with the sum representation of the kernel~\eqref{eq:main1a}. The actions can be approximated then by
\begin{equation}
\begin{split}\label{saddle-expansion.d}
-\mathcal{S}(j;x)+\mathcal{S}(t;y)\overset{M,N\gg1}{\approx}&-\mathcal{S}(\widehat{x};x)+\mathcal{S}(\widehat{y};y)-\frac{(M+1)\psi'(1+\widehat{x})}{2}(j-\widehat{x})^2-\frac{(M+1)\psi'(1+\widehat{y})}{2}\delta t^2-i\pi\sign(\delta t)
\end{split}
\end{equation}
with $t=\widehat{y}+i\delta t$. Plugging this into the kernel~\eqref{eq:main1a}, we obtain
\begin{equation}
\begin{split}
K_Y(x,y)\frac{dy}{d\widehat{y}} =& (M+1)\psi'(1+\widehat{y})\exp\left[-\mathcal{S}(\widehat{x};x)+\mathcal{S}(\widehat{y};y)\right]\\
&\times\sum_{j=0}^{N-1} \int_{-\infty}^\infty \frac{dt}{2\pi } \frac{\sin (\pi (j-\widehat{y}-i\delta t))}{\pi (j-\widehat{y}-i\delta t)} \exp\left[-\frac{(M+1)\psi'(1+\widehat{x})}{2}(j-\widehat{x})^2-\frac{(M+1)\psi'(1+\widehat{y})}{2}\delta t^2\right].
\end{split}
\end{equation}
The Gaussian can be anew replaced by Dirac delta functions since their variance shrinks like $1/M$. Moreover, we can take the limit of the upper boundary of $j$ {to infinity} because it has to be close to $\widehat{x}$ which is of order one. Multiplying the kernel with the factor $\sqrt{\psi'(1+\widehat{x})/\psi'(1+\widehat{y})}\exp\left[\mathcal{S}(\widehat{x};x)-\mathcal{S}(\widehat{y};y)\right]$, see \eqref{gauge}, leads us to the final result of this section
\begin{equation}\label{pf-hard}
K_{\rm pf}^+(\widehat{x},\widehat{y})
=\lim_{M,N\to\infty}\sqrt{\frac{\psi'(1+\widehat{x})}{\psi'(1+\widehat{y})}}\exp\left[\mathcal{S}(\widehat{x};x)-\mathcal{S}(\widehat{y};y)\right]K_Y(x,y)\frac{dy}{d\widehat{y}} =\sum_{j=0}^{\infty} \frac{\sin [\pi (j-\widehat{y})]}{\pi (j-\widehat{y})} \delta(j-\widehat{x}).
\end{equation}
This kernel 
{agrees with}
~\eqref{pf-bulk} -- apart from the shift $\nu$ -- and, indeed, it is the spectral statistics of a picket fence spectrum with a lower bound, {whence the superscript $+$}. The sine function again ensures that not two eigenvalues sit at the same position.

We would like to emphasise that this part of the spectrum always shows up regardless how $M$ and $N$ are sent to infinity as only the relation between $\RE[z_{\rm s}]$ and $M$ have been important for the derivation.

\section{Soft Edge}\label{sec:softedge}

We start again from~\eqref{eq:main1a} and exploit the knowledge that the action is convex on the interval $]-1,N[$. At the soft edge the spectral variable  $y$ grows so  {strongly} in the large $N,M$ limit, see Sec.~\ref{sec:bulk}, that $z_0$, which is defined by~\eqref{saddle-real}, is close to the value $N-1$, in particular $N-1-z_0=\mathcal{O}(1)$. The question is again how many summands can contribute. To solve this problem we consider the scaling bound {on $\Delta\leq\mathcal{O}(1)$~\eqref{rec:order} defined in \eqref{difference.b}.  It has been obtained from the difference of the action at $j$ and $z_0$ which has to be of order one or less to be contributing, as $\Delta\gg1$ will be exponentially suppressed in the sum.} In the present case we have to replace $z_0$ by $N-1-\widehat{z}_0$ with $\widehat{z}_0=\mathcal{O}(1)$ such that we consider
\begin{equation}\label{rec:order.b}
|\delta j|\leq\mathcal{O}\left(\frac{1}{\sqrt{\int_0^1d\lambda(1-\lambda)\left[(M+1)\psi'(N-\widehat{z}_0+\delta j\lambda)+\psi'(\widehat{z}_0+1-\delta j\lambda)\right]}}\right)
\end{equation}
with $\delta j= j-N+1+\widehat{z}_0$. When assuming that $|\delta j|<\mathcal{O}(N)$ and plugging this into~\eqref{rec:order.b}, we get the new tighter bound
\begin{equation}\label{rec:order.c}
|\delta j|\leq\mathcal{O}\left(\min\left\{\sqrt{\frac{N}{M}},\sqrt{|\delta j|}\right\}\right).
\end{equation}
We can combine this inequality with the knowledge that the smallest non-vanishing $|\delta j|$  is equal to $1$. This leads to $|\delta j|\leq\mathcal{O}(1)$ which is immediate for $N\leq M$. For $N\gg M$, we can plug $|\delta j|\leq \mathcal{O}(\sqrt{N/M)}$ back into~\eqref{rec:order.c} which shows $|\delta j|\leq\mathcal{O}\left(\sqrt{|\delta j|}\right)$ and, thus, $|\delta j|\leq\mathcal{O}(1)$.

For the integration variable $t=z_0+i\delta t$, we are looking for the maximum $t_{\max}$ of $\RE[S(z_0+i\delta t,y)]$ with respect to $\delta t$. Its derivative yields the imaginary part of the saddle point equation~\eqref{saddlepointeqbulk} with the fixed real part $\RE(z_s)\to z_0$. The upper bound~\eqref{imaginary-bounds} for the positions of the maxima still holds so that we have still the scaling bound $t_{\max}\leq\mathcal{O}(z_0/M)=\mathcal{O}(N/M)$. The lower bound in~\eqref{imaginary-bounds} is not useful anymore because $z_0$ can be very close to $N-1$.

Let us point out that $\Im[S'(z_0+i\delta t,y)]$ is strictly decreasing for $\delta t\in\mathbb{R}_-$ and for $\delta t\in\mathbb{R}_+$, separately. Thus, $\RE[S(z_0+i\delta t,y)]$ is concave  on both of the two half axes $\mathbb{R}_-$ and $\mathbb{R}_+$ so that the maxima are unique on each of the two parts. What we have to check is the width of the maximum which is given by the second derivative of $\RE[S(z_0+i\delta t,y)]$ with respect to $\delta t$ at $t_{\max}$ 
\begin{equation}
-\RE[S''(z_0+it_{\max},y)]=-(M+1)\RE[\psi'(1+z_0+it_{\max})]+\RE[\psi'(N-z_0-it_{\max})]=\mathcal{O}\left(\max\left\{\frac{M}{N},1\right\}\right),
\end{equation}
where we have used $t_{\max}\leq\mathcal{O}(N/M)$ and the asymptotic expansion~\eqref{asymptotic.Digamma}. Hence, when expanding about $\delta t=0$ it is safe to say that we choose $\delta t$ of order $\mathcal{O}(\max\{N/M,1\})$. It is very important that we take $N/M$ and not the width of the maximum $\sqrt{N/M}$ since for the case $N\gg M$ the maximum might be further away from the real axis than the width of the maximum is covering.

In summary, we choose $j=N-1-\widehat{j}$ with $\widehat{j}$ being of order one and also shift the integration variable as follows $t=N-1-\widehat{j} +i\widehat{t}$. Then, the kernel becomes
\begin{equation}
\begin{split}
\left(\frac{y}{x}\right)^{N-1}K_Y(x,y) =& \frac{1}{y}\sum_{\widehat{j}=0}^{N-1} \int_{-\infty}^\infty \frac{d\widehat{t}}{2\pi } \frac{\sinh [\pi \widehat{t}]}{\pi \widehat{t}}\frac{x^{-\widehat{j}}}{y^{i\widehat{t}-\widehat{j}}} \left(\frac{\Gamma[N-\widehat{j}+i\widehat{t}]}{(N-1-\widehat{j})!}\right)^{M+1}\frac{\Gamma[1+\widehat{j}-i\widehat{t}]}{\widehat{j}!}\\
=&\frac{1}{y}\sum_{\widehat{j}=0}^{N-1} \int_{-\infty}^\infty \frac{d\widehat{t}}{2\pi i\widehat{t}y^{i\widehat{t}}}  \left(\frac{\Gamma[N+i\widehat{t}]}{(N-1)!}\exp\left[-\sum_{l=1}^{\widehat{j}}{\rm ln}\left(1+i\frac{\widehat{t}}{N-l}\right)\right]\right)^{M+1}\frac{\left(-y/x\right)^{\widehat{j}}}{\widehat{j}!\Gamma[i\widehat{t}-\widehat{j}]}.
\end{split}
\end{equation}
In the second line we have exploited Euler's reflection formula~\eqref{reflect} for $\Gamma[1+\widehat{j}-i\widehat{t}]$. {We will remove a factor $(y/x)^{N-1}$ from this intermediate result by the invariance \eqref{gauge} out of convenience as we will see below.}

In the next step, we can approximate the exponential
\begin{equation}
\exp\left[-(M+1)\sum_{l=1}^{\widehat{j}}{\rm ln}\left(1+i\frac{\widehat{t}}{N-l}\right)\right]\overset{M,N}{\approx}\exp\left[-i(M+1)\frac{\widehat{t}}{N}\widehat{j}\right]
\end{equation}
because $\widehat{t}$ is of order $N/M$ or smaller so that all higher order terms in the expansion above vanish in the limit regardless how $M$ and $N$ are related. Afterwards, the sum over $\widehat{j}$ can be extended to a series over $\mathbb{N}_0$ since all terms which are not of order one will be suppressed in the large $N,M$ limit as we have noticed in the previous discussion. The benefit of this extension is the binomial series which can be carried out now, i.e.
\begin{equation}\label{binomial.trick}
\sum_{\widehat{j}=0}^{N-1}\frac{\left(-e^{-i(M+1)\widehat{t}/N}y/x\right)^{\widehat{j}}}{\widehat{j}!\Gamma[i\widehat{t}-\widehat{j}]}\approx\sum_{\widehat{j}=0}^{\infty}\frac{\left(-e^{-i(M+1)\widehat{t}/N}y/x\right)^{\widehat{j}}}{\widehat{j}!\Gamma[i\widehat{t}-\widehat{j}]}=\frac{1}{\Gamma[i\widehat{t}]}\left(1-\frac{y}{x}e^{-i(M+1)\widehat{t}/N}\right)^{i\widehat{t}-1}.
\end{equation}

We eventually arrive at the following intermediate result
\begin{equation}\label{kernel.intermediate}
\begin{split}
\left(\frac{y}{x}\right)^{N-1}K_Y(x,y) \overset{M,N\gg1}{\approx}& \frac{1}{y}\int_{-\infty}^\infty \frac{d\widehat{t}}{2\pi }\frac{y^{-i\widehat{t}}}{\Gamma[1+i\widehat{t}]} \left(\frac{\Gamma[N+i\widehat{t}]}{(N-1)!}\right)^{M+1}\left(1-\frac{y}{x}e^{-i(M+1)\widehat{t}/N}\right)^{i\widehat{t}-1}.
\end{split}
\end{equation}
This representation is ideal to perform the three different double scaling limits in the ensuing subsections.

%%%%%%%%%%%%%%%%%%%%%%%%%%%%%%%%%%%%%
\subsection{Airy-kernel -- $\bf N\gg M$}\label{sec:Airy}

As we have seen for the double scaling $N\gg M$, we may have to go in the integration over $\widehat{t}$ up to the scale $N/M$. This is certainly a rough estimate, but it at least covers the contributing parts of the integrand. Indeed, we will see that the scale where $\widehat{t}$ is contributing is actually smaller.

{In subsection~\ref{sub:meso}} we have seen that the soft edge lies at $y_0=y_+=e (M+1)N^M$. Moreover, it vanishes like the square root~\eqref{density-near-soft}, so that the local scale is additionally multiplied by $(M/N)^{2/3}$. Therefore, we consider the change of variables
\begin{equation}\label{non-unfold-soft-airy}
x=e (M+1)N^Me^{(M/N)^{2/3}\chi}\ {\rm and}\ y=e (M+1)N^Me^{(M/N)^{2/3}\zeta}\ {\rm with}\ \chi,\eta=\mathcal{O}(1).
\end{equation}
The coordinates $\chi$ and $\zeta$ are not yet properly unfolded, and they will need to be corrected for the square root behaviour. The kernel will, however, exhibit the well-known form of the Airy-kernel~\cite{Peter}.
To see this we expand the logarithm of the integrand of~\eqref{kernel.intermediate} about $\widehat{t}=0$, apart from $\Gamma[1+i\widehat{t}]$ where the argument $\widehat{t}$ is large compared to $1$, so that we use Stirlings formula for this term. 
In particular we employ the expansions
\begin{equation}\label{expan.a}
\begin{split}
-{\rm ln}(\Gamma[1+i\widehat{t}])=&-i\widehat{t}\,{\rm ln}(i\widehat{t})+i\widehat{t}-\frac{1}{2}{\rm ln}(i\widehat{t})-\frac{{\rm ln}(2\pi)}{2}+\mathcal{O}\left(\frac{1}{|\widehat{t}|}\right),\\
(M+1){\rm ln}\left(\frac{\Gamma[N+i\widehat{t}]}{(N-1)!}\right)=&(M+1)\left(i{\rm ln}(N)\widehat{t}-\frac{1}{2N}\widehat{t}^2\right)+\mathcal{O}\left(\frac{M|\widehat{t}|}{N}+\frac{M|\widehat{t}|^2}{N^2}+\frac{M|\widehat{t}|^3}{N^2}\right),
\end{split}
\end{equation}
and
\begin{equation}\label{expan.b}
\begin{split}
&(i\widehat{t}-1){\rm ln}\left(1-e^{(M/N)^{2/3}(\zeta-\chi)-i(M+1)\widehat{t}/N}\right)\\
=&(i\widehat{t}-1)\left[\left(\frac{M}{N}\right)^{2/3}(\zeta-\chi)+{\rm ln}\left(1-e^{-i(M+1)\widehat{t}/N}\right)+{\rm ln}\left(1+\frac{e^{-(M/N)^{2/3}(\zeta-\chi)}-1}{1-e^{-i(M+1)\widehat{t}/N}}\right)\right]\\
=&(i\widehat{t}-1){\rm ln}\left(i\frac{M+1}{N}\widehat{t}\right)+\frac{1}{2}\frac{M}{N}\widehat{t}^2-\frac{i}{24}\left(\frac{M}{N}\right)^2\widehat{t}^3+i\frac{1}{2}\left(\frac{M}{N}\right)^{2/3}(\zeta-\chi)\widehat{t}+\frac{i}{2}\left(\frac{N}{M}\right)^{2/3}\frac{(\zeta-\chi)^2}{\widehat{t}}\\
&-\left(\frac{N}{M}\right)^{1/3}(\zeta-\chi)+\mathcal{O}\left[\frac{M}{N}|\widehat{t}|+\left(\frac{N}{M}\right)^{1/3}\frac{1}{|\widehat{t}|}+\left(\frac{M}{N}\right)^{5/3}|\widehat{t}|^2+\left(\frac{M}{N}\right)^3|\widehat{t}|^4+\frac{N}{M}\frac{1}{|\widehat{t}|^2}\right].
\end{split}
\end{equation}
Combining these three terms with $y^{-i\widehat{t}}$. The kernel takes the form
\begin{equation}
\begin{split}
&\frac{y^{N-1}\exp\left[(N/M)^{1/3}\zeta\right]}{x^{N-1}\exp\left[(N/M)^{1/3}\chi\right]}K_Y(x,y)\frac{dy}{d\zeta}\\
=&\left(\frac{N}{M}\right)^{1/3}\int_{-\infty}^\infty \frac{d\widehat{t}}{(2\pi i\widehat{t})^{3/2}}\exp\left[-\frac{i}{24}\left(\frac{M}{N}\right)^2\widehat{t}^3-\frac{i}{2}\left(\frac{M}{N}\right)^{2/3}(\zeta+\chi)\widehat{t}+\frac{i}{2}\left(\frac{N}{M}\right)^{2/3}\frac{(\zeta-\chi)^2}{\widehat{t}}\right]\\
&\times\exp\left[\mathcal{O}\left(\frac{M}{N}|\widehat{t}|+\left(\frac{N}{M}\right)^{1/3}\frac{1}{|\widehat{t}|}+\left(\frac{M}{N}\right)^{5/3}|\widehat{t}|^2+\left(\frac{M}{N}\right)^3|\widehat{t}|^4+\frac{|\widehat{t}|^2}{N}\right)\right].
\end{split}
\end{equation}
When  $\widehat{t}$ scales like $(N/M)^{2/3}$ the first part of the exponential is of order one while the corrections are all smaller than one and, hence, vanish in the large $N,M$ limit. This is the aforementioned scaling that is smaller than the crude approximation by $N/M$. Therefore, we choose $\widehat{t}=(N/M)^{2/3}\delta t$ with $\delta t=\mathcal{O}(1)$.

Finally, we arrive at the limiting soft edge kernel for $N\gg M$ which is the Airy-kernel
\begin{equation}\label{airy-non-unfold}
\begin{split}
K_{\rm Airy}(\chi,\zeta)=&\lim_{N,M\to\infty}\frac{y^{N-1}\exp\left[(N/M)^{1/3}\zeta\right]}{x^{N-1}\exp\left[(N/M)^{1/3}\chi\right]}K_Y(x,y)\frac{dy}{d\zeta}\\
=&\int_{-\infty}^\infty\frac{d\delta t}{(2\pi(c+ i \delta t))^{3/2}}\exp\left[\frac{1}{24}(c+i\delta t)^3-\frac{1}{2}(\zeta+\chi)(c+i\delta t)-\frac{1}{2}\frac{(\zeta-\chi)^2}{c+i\delta t}\right],
\end{split}
\end{equation}
where $c>0$ is a positive shift to guarantee the convergence. The integration can be made absolutely integrable by 
{tilting} the two half axes.

The integral above is none of the standard representations of the Airy-kernel. That can be obtained by introducing a Gaussian integral {for the last term in the exponential. After additionally rescaling $c+i\delta t \to 2(c+i\delta t)$ we have}
\begin{equation}\label{airy-non-unfold.b}
\begin{split}
K_{\rm Airy}(\chi,\zeta)=&
\int_{-\infty}^\infty \frac{d\delta t}{2\pi(c+i\delta t)}\int_{-\infty}^\infty \frac{d\delta s}{2\pi} \exp\left[\frac{(c+i\delta t)^3}{6}-\zeta(c+i\delta t)+(c+i\delta t)(c-i\delta s)^2-(\chi-\zeta)(c-i\delta s)\right]
\\
=&
\int_{-\infty}^\infty \frac{d\delta t}{2\pi(c+i\delta t)}\int_{-\infty}^\infty \frac{d\delta s}{2\pi} \exp\left[\frac{(c+i\delta t)^3}{6}+\frac{(c-i\delta s)^3}{6}-\zeta(c+i\delta t)-\chi(c-i\delta s)\right]
\\
=&\ 
2^{1/3}\int_0^\infty dr {\rm Ai}[2^{1/3}(\zeta+r)]\,{\rm Ai}[2^{1/3}(\chi+r)].
\end{split}
\end{equation}
In the second step we have shifted, first, $c+i\delta t \to (2c+i\delta t-i\delta s)$, and {in the final one we have introduced the integral
$1/(2c+i\delta t-i\delta s)=\int_0^\infty dr \exp[-(2c+i\delta t-i\delta s)r]$. Moreover, we have
rescaled $(c+i\delta t)\to 2^{1/3}(c+i\delta t)$ and $(c-i\delta s)\to 2^{1/3}(c-i\delta s)$ and used the integral representation~\cite{NIST} of the  Airy function ${\rm Ai}$. The final line of~\eqref{airy-non-unfold.b} is one of the common representations of the Airy-kernel.}

As already mentioned, the coordinates~\eqref{non-unfold-soft-airy} are not properly unfolded. When going into the bulk, the level spacing will shrink, due to the square root increase. This is quite unfortunate while connecting these statistics with the uniform picket fence spectrum, which we will derive in the next subsection for the opposite scaling relation $M\gg N$. To amend this problem, we change variables,
\begin{equation}\label{unfold-soft-airy}
x=e (M+1)N^M\left(1+\sign(\widehat{x})\left[\frac{3 \pi (M+1)}{\sqrt{8}N}|\widehat{x}|\right]^{2/3}\right)\ {\rm and}\ y=e (M+1)N^M\left(1+\sign(\widehat{y})\left[\frac{3 \pi (M+1)}{\sqrt{8}N}|\widehat{y}|\right]^{2/3}\right)
\end{equation}
with $\widehat{x},\widehat{y}=\mathcal{O}(1)$ instead of~\eqref{non-unfold-soft-airy}, that follows from the mesoscopic level density~\eqref{density-near-soft} very close to the soft edge. The unfolding outside the support of the mesoscopic level density is {somewhat} 
artificial, but in this way one can still catch the tail of the largest eigenvalue. There is certainly one disadvantage, that we get a coordinate singularity at the origin, which is chosen to be the edge of the mesoscopic level density.

The kernel in the new coordinates can be readily obtained by noticing that
\begin{equation}
\zeta=\left(\frac{N}{M}\right)^{2/3}{\rm ln}\left(1+\sign(\widehat{y})\left[\frac{3\pi (M+1)}{\sqrt{8}N}|\widehat{y}|\right]^{2/3}\right)\approx\sign(\widehat{y})\frac{(3\pi|\widehat{y}|)^{2/3} }{2}
\end{equation}
and similarly for the relation between $\widehat{x}$ and $\chi$. Hence, we have {after unfolding}
\begin{equation}\label{airy-unfold}
\begin{split}
&\widehat{K}_{\rm Airy}(\widehat{x},\widehat{y})=K_{\rm Airy}(\chi,\zeta)\frac{d\zeta}{d\widehat{y}}\\
=&\frac{2^{2/3}}{3|\widehat{y}|^{1/3}}\frac{{\rm Ai}(\sign(\widehat{x})(3\pi|\widehat{x}|/2)^{2/3}){\rm Ai}'(\sign(\widehat{y})(3\pi|\widehat{y}|/2)^{2/3})-{\rm Ai}'(\sign(\widehat{x})(3\pi|\widehat{x}|/2)^{2/3}){\rm Ai}(\sign(\widehat{y})(3\pi|\widehat{y}|/2)^{2/3})}{\sign(\widehat{x})|\widehat{x}|^{2/3}-\sign(\widehat{y})|\widehat{y}|^{2/3}}.
\end{split}
\end{equation}
The limit $\widehat{x}\to\widehat{y}$ yields the unfolded microscopic level density which is equal to
\begin{equation}\label{airy-density-unfold}
\widehat{\rho}_{\rm Airy}(\widehat{y})=\left(\frac{2\pi^2}{3|\widehat{y}|}\right)^{1/3}\left(\left[{\rm Ai}'\left(\sign(\widehat{y})\left(\frac{3\pi}{2}|\widehat{y}|\right)^{2/3}\right)\right]^2-\sign(\widehat{y})\left(\frac{3\pi}{2}|\widehat{y}|\right)^{2/3}\left[{\rm Ai}\left(\sign(\widehat{y})\left(\frac{3\pi}{2}|\widehat{y}|\right)^{2/3}\right)\right]^2\right).
\end{equation}
As can be easily checked, the local mean level spacing equals one even for the  
{largest eigenvalues (mapped to the origin here).}

%%%%%%%%%%%%%%%%%%%%%%%%%%%%%%%%%
\subsection{Picket Fence at the Soft Edge -- $\bf N\ll M$}\label{sec:pf.soft}

For the other extreme case $M\gg N$, the maxima of the action are very close to the origin, {where the soft edge is located}. The properly unfolded scaling variables are
\begin{equation}\label{unfolding.llM.soft}
x=\exp\left[(M+1)\psi(N)+M\psi'(N)\widehat x\right]\ {\rm and}\ y=\exp\left[(M+1)\psi(N)+M\psi'(N)\widehat y\right]\ {\rm with}\ \widehat x,\widehat y=\mathcal{O}(1),
\end{equation}
in particular it is $z_0\approx N-1$. This unfolding is essentially the same as in~{\eqref{unfolding.llM} } up to the scaling of $z_0$. This is the reason why we can still use the intermediate result~\eqref{pf-kernel-interp}, because the arguments still hold. The only difference is that $j=N-1-\widehat{j}$ has to be in the vicinity of $N-1$, i.e., $\widehat{j}=\mathcal{O}(1)$. This means that we can extend the sum over $\widehat{j}$ 
{from $-\infty$ to $0$.}
This leads to the result
\begin{equation}\label{pf-soft}
\begin{split}
K_{\rm pf}^-(\widehat{x},\widehat{y})
=&\lim_{M,N\to\infty} \exp\left[-\frac{M\psi'(N)}{2}([\widehat{x}+N-1]^2-[\widehat{y}+N-1]^2)\right]K_Y(x,y) \frac{dy}{d\widehat{y}}\\
=&\sum_{\widehat{j}=-\infty}^{0}\frac{\sin (\pi (-\widehat{j}+\widehat{y}))}{\pi (-\widehat{j}+\widehat{y})}\delta(-\widehat{j}+\widehat{x})\\
=&K_{\rm pf}^+(-\widehat{x},-\widehat{y}).
\end{split}
\end{equation}
We see that 
up to a reflection 
{it agrees with the one at the hard edge \eqref{pf-hard}}. It is not very surprising since we are at the upper 
{edge} 
of the picket fence spectrum. Therefore, it is also properly unfolded implying the mean level spacing is one.

%%%%%%%%%%%%%%%%%%%%%%%%%%%%%%%
\subsection{Critical Regime at the Soft Edge -- $\bf N\propto M$}\label{sec:trans.soft}

In the critical regime when $M=\mathcal{O}(N)$, the integration variable in $\widehat{t}$ in~\eqref{kernel.intermediate} is of order one. Thence, we only need to expand the term ${\rm ln}(\Gamma[N+i\widehat{t}])$ in $\widehat{t}$ and choose the variables
\begin{equation}\label{crit-soft-non-unfold}
x=N^{M+1}e^{\chi}\ {\rm and}\ y=N^{M+1}e^{\zeta}\ {\rm with}\ \chi,\zeta=\mathcal{O}(1).
\end{equation}
We would like to mention that the new variables $\chi$ and $\zeta$ are again not properly unfolded, yet, since they show a transition between the uniformly distributed picket fence spectrum at $M\gg N$ and the square root behaviour for $N\gg M$, see Sec.~\ref{sec:bulk}.
The kernel in this {critical regime at the soft edge (cs)} is then
\begin{equation}\label{kernel.critical.a}
\begin{split}
K_{\rm cs}(\chi,\zeta)=&\lim_{\substack{M,N\to\infty\\N/M\to a}}\left(\frac{y}{x}\right)^{N-1}K_Y(x,y)\frac{dy}{d\zeta}\\
=&\int_{-\infty}^\infty \frac{d\widehat{t}}{2\pi }\frac{\left(1-\exp[\zeta-\chi-i\widehat{t}/a]\right)^{i\widehat{t}-1}}{\Gamma[1+i\widehat{t}]}\exp\left[-\frac{1}{2a}\widehat{t}^2-i\left(\zeta+\frac{1}{2a}\right) \widehat{t}\right].
\end{split}
\end{equation}
We have only approximated $(\Gamma[N+i\widehat{t}]/\Gamma[N])^{M+1}\approx N^{i(M+1)\widehat{t}}\exp[-(M+1)\widehat{t}^2/(2N)-iM\widehat{t}/(2N)]$ {and $a=N/M$.}
Apart from some rescaling this is the result in~\cite{ABK}.

Another representation of this kernel has been derived in~{\cite{LWW}} and has the form of a double contour integral
\begin{equation}\label{kernel.critical.b}
K_{\rm cs}(\chi,\zeta)=\oint_{\mathcal{C}}\frac{ds}{2\pi i}\int_{-\infty}^\infty\frac{dt}{2\pi}\frac{1}{1+it-s}\frac{\Gamma[s]}{\Gamma[1+it]}\frac{\exp[1/(2a)(1+it)^2-(\zeta+1/(2a)) (1+it)]}{\exp[1/(2a)s^2-(\chi+1/(2a)) s]},
\end{equation}
where {the contour $\mathcal{C}$ encircles the poles at $s=0,-1,-2,\ldots$ of $\Gamma[s]$}. One can readily show that both expressions agree by taking the residues at $s=0,-1,-2\ldots$, yielding
\begin{equation}
K_{\rm cs}(\chi,\zeta)=\sum_{\widehat{j}=0}^\infty\int_{-\infty}^\infty\frac{dt}{2\pi}\frac{1}{1+it+\widehat{j}}\,\frac{(-1)^{\widehat{j}}}{\widehat{j}!\Gamma[1+it]}\frac{\exp[1/(2a)(1+it)^2-(\zeta+1/(2a)) (1+it)]}{\exp[1/(2a)\widehat{j}^2+(\chi+1/(2a)) \widehat{j}]}.
\end{equation}
Next, we shift $it=i\widehat{t}-\widehat{j}-1$ and perform the resulting binomial series with the help of~\eqref{binomial.trick}. Then we arrive at~\eqref{kernel.critical.a}.

Let us see how we get the other scaling limits from Subsections~\ref{sec:Airy} and~\ref{sec:pf.soft}.
When we rescale $\chi=M\widehat{x}/N$ and $\zeta=M\widehat{y}/N$ and take the limit $N/M=a\to0$, we can regain the result~\eqref{pf-soft}. {To this aim, we expand the bracket in~\eqref{kernel.critical.a} via reading~\eqref{binomial.trick} backwards and then shift $\widehat{t}\to\widehat{t}-i(\widehat{j}+\widehat{y}+1/2)$. This leads to four Gaussian terms,
\begin{equation}
\begin{split}
\frac{M}{N}K_{\rm cs}\left(\frac{M}{N}\widehat{x},\frac{M}{N}\widehat{y}\right)=&\frac{M}{N}
{\sum_{\widehat{j}=0}^\infty}
\int_{-\infty}^\infty \frac{d\widehat{t}}{2\pi }(-1)^{\widehat{j}}\frac{\exp\left[-\frac{\widehat{t}^2}{2}-\frac{M}{N}\widehat{j}(\widehat{x}-\widehat{y})-\frac{M}{2N}\left(\widehat{j}+\widehat{y}+\frac{1}{2}\right)^2\right]}{(\widehat{j}+\widehat{y}+1/2+i\widehat{t})\widehat{j}!\Gamma[\widehat{y}+1/2+i\widehat{t}]}\\
\approx&\frac{M}{N}
{\sum_{\widehat{j}=0}^\infty}
\int_{-\infty}^\infty \frac{d\widehat{t}}{2\pi }(-1)^{\widehat{j}}\frac{\exp\left[-\frac{M}{2N}\widehat{t}^2-\frac{M}{2N}\left(\widehat{j}+\widehat{x}+\frac{1}{2}\right)^2+\frac{M}{2N}\left(\widehat{x}+\frac{1}{2}\right)^2-\frac{M}{2N}\left(\widehat{y}+\frac{1}{2}\right)^2\right]}{(\widehat{j}+\widehat{y}+1/2+i\widehat{t})\widehat{j}!\Gamma[\widehat{y}+1/2+i\widehat{t}]}.
\end{split}
\end{equation}
The Gaussians for $\widehat{t}$ and $\widehat{j}$ can be replaced by Dirac delta functions as their variance shrinks with $N/M\to0$,
\begin{equation}
\begin{split}
\frac{M}{N}K_{\rm cs}\left(\frac{M}{N}\widehat{x},\frac{M}{N}\widehat{y}\right)\approx&\exp\left[\frac{M}{2N}\left(\widehat{x}+\frac{1}{2}\right)^2-\frac{M}{2N}\left(\widehat{y}+\frac{1}{2}\right)^2\right]{\sum_{\widehat{j}=0}^\infty}
(-1)^{\widehat{j}}\frac{\delta(\widehat{j}+\widehat{x}+1/2)}{(\widehat{j}+\widehat{y}+1/2)\widehat{j}!\Gamma[\widehat{y}+1/2]}.
\end{split}
\end{equation}
For the Gamma function $\Gamma[\widehat{y}+1/2]$ we apply Euler's reflection formula~\eqref{reflect} and combine it with the sign $(-1)^{\widehat{j}}$. Additionally, we can replace $\widehat{j}!$ by $\Gamma[1/2-\widehat{x}]$, which leads us to the final result
\begin{equation}
\lim_{\substack{M/N\to\infty}}\frac{M}{N}\frac{\Gamma[1/2-\widehat{x}]\exp\left[-\frac{M}{2N}\left(\widehat{x}+\frac{1}{2}\right)^2\right]}{\Gamma[1/2-\widehat{y}]\exp\left[-\frac{M}{2N}\left(\widehat{y}+\frac{1}{2}\right)^2\right]}K_{\rm cs}\left(\frac{M}{N}\widehat{x},\frac{M}{N}\widehat{y}\right)=K_{\rm pf}^-\left(\widehat{x}+\frac{1}{2},\widehat{y}+\frac{1}{2}\right).
\end{equation}
The shift by $1/2$ results from the fact that the square root behaviour in the other double scaling limits pushes the largest eigenvalue slightly into the bulk, away from the edge of the mesoscopic support of the level density.
}

In the opposite limit $N/M=a\to\infty$, we choose
\begin{equation}
\chi=1-{\rm ln}\left(\frac{N}{M}\right)+\left(\frac{M}{N}\right)^{2/3}\chi'\quad{\rm and}\quad \zeta=1-{\rm ln}\left(\frac{N}{M}\right)+\left(\frac{M}{N}\right)^{2/3}\zeta'.
\end{equation}
We need to expand only the ratio in the integral~\eqref{kernel.critical.a} by using the expansions in~\eqref{expan.a} and~\eqref{expan.b}. This directly leads to the limit
\begin{equation}
\lim_{N/M\to\infty}\left(\frac{M}{N}\right)^{2/3}\frac{\exp[-(\frac{N}{M})^{1/3}\zeta']}{\exp[-(\frac{N}{M})^{1/3}\chi']}K_{\rm cs}\left(1-{\rm ln}\left(\frac{N}{M}\right)+\left(\frac{M}{N}\right)^{2/3}\chi',1-{\rm ln}\left(\frac{N}{M}\right)+\left(\frac{M}{N}\right)^{2/3}\zeta'\right)=K_{\rm Airy}(\chi',\zeta').
\end{equation}

\begin{figure}[t!]
 \centering
\includegraphics[width=0.8\textwidth]{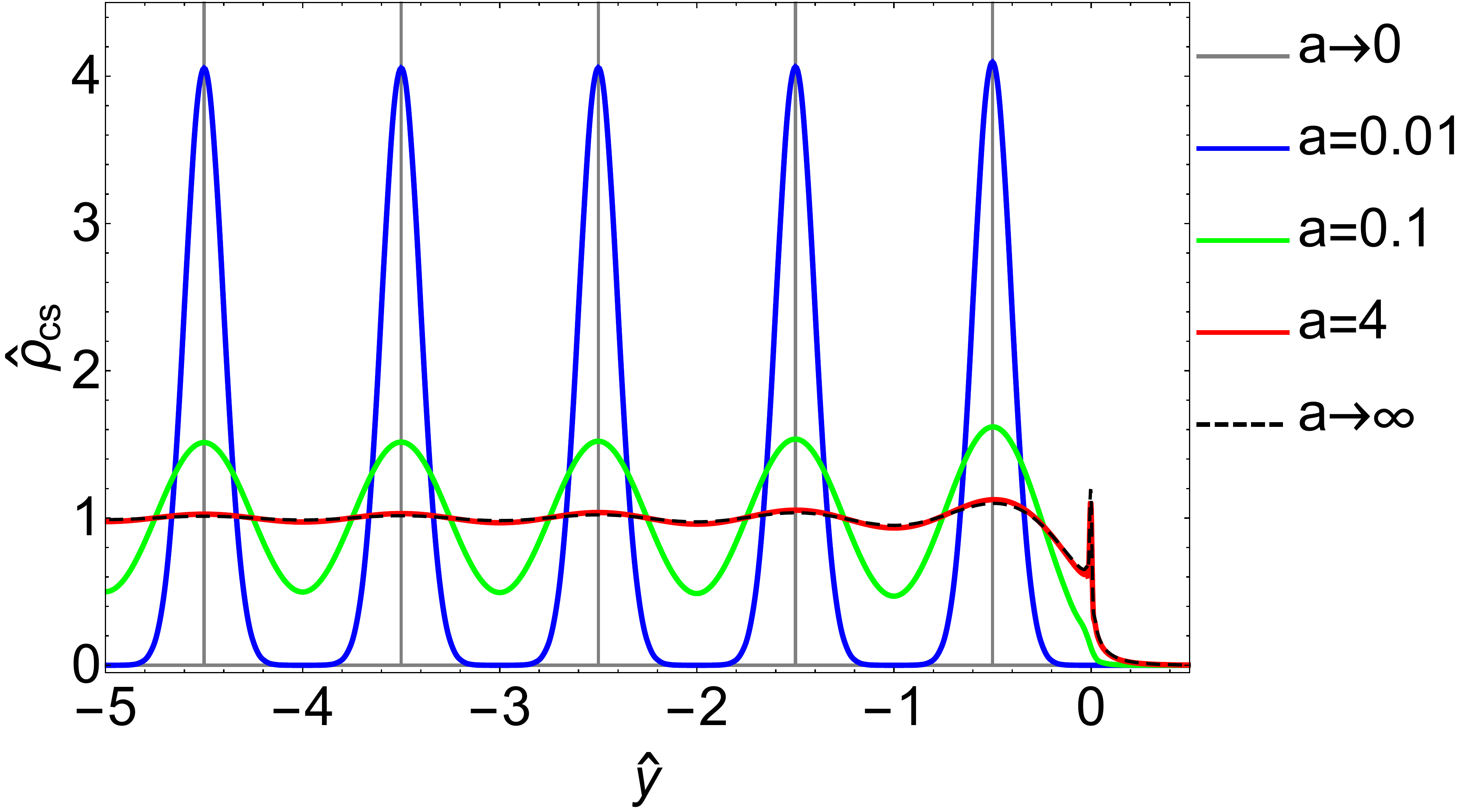}
 \caption{The properly unfolded level density~\eqref{crit-soft-unfold} at the soft edge for various ratios $a=N/M$. The bulk of eigenvalues is to the left. The peak of the distributions at the origin is the price we have to pay of unfolding which generates a coordinate singularity, since the mesoscopic level density vanishes there. Note that the individual eigenvalues are shifted by $1/2$ and do not lie on the integers.}
 \label{fig:unfolddensity}
\end{figure}

As both limits need a rescaling, we immediately notice that the choice~\eqref{crit-soft-non-unfold} has not been the proper unfolding otherwise the mean level spacing would have been the same. As for the Airy-kernel in subsection~\ref{sec:Airy}, we employ the unfolding {on} the mesoscopic scale Eq.~\eqref{unfold.sb}. Despite the fact that it has been derived for $N\gg M$ it still works out for $N=\mathcal{O}(M)$, because the number of eigenvalues that sit in
{the tail which is  suppressed by $1/M$},  is of order $N/M=\mathcal{O}(1)$. In Fig.~\ref{fig:unfolddensity}, we show that the scaling
\begin{equation}\label{unfold.sb.b}
\begin{split}
\phi(\widehat{y})=&\int_{-\pi}^\pi\Theta\left[\widehat{y}-\frac{N}{M\pi}\left(\phi'-\frac{{\phi'}^2}{\tan(\phi')}\right)\right] d\phi'-\pi\  {\rm with}\  \zeta={\rm ln}\left(\frac{M}{N}\right)+1+\sign(\widehat{y})\left[{\rm ln}\left(\frac{\sin[\phi(\widehat{y})]}{\phi(\widehat{y})}\right)+\frac{\phi(\widehat{y})}{\tan[\phi(\widehat{y})]}-1\right].
 \end{split}
\end{equation}
yields indeed a properly unfolded microscopic level density
\begin{equation}\label{crit-soft-unfold}
\widehat{\rho}_{\rm cs}(\widehat{y})=K_{\rm cs}(\zeta,\zeta)\left|\frac{d\zeta}{d\widehat{y}}\right|=\frac{M\pi}{N|\phi(\zeta(\widehat{y}))|}K_{\rm cs}(\zeta(\widehat{y}),\zeta(\widehat{y})).
\end{equation}
In this figure, we notice that the true soft edge is slightly shifted inside {by $1/2$,} when going from the Airy statistics to the picket  fence statistics. The deeper reason for this is that the width of  the distributions of the individual eigenvalues becomes increasingly narrow, and the mean of the largest eigenvalue was always inside the bulk of the spectrum. This is a rather fascinating effect, which is shared with a non-zero vacuum energy of the Harmonic oscillator. The question is whether there is some deeper physical meaning behind this observation.

%%%%%%%%%%%%%%%%%%%%%%%%%%%%%%%%%%%
\section{Duality between Level Statistics for Products and Brownian Motion} 
\label{sec:duality}

The evolution of eigenvalues of a Gaussian matrix 
whose elements perform independent Brownian motions is known as 
Dysonian Brownian Motion. The problem of Dysonian Brownian Motion 
with the initial condition given by equidistant eigenvalues was solved
in \cite{KurtBrown}. The statistics of eigenvalues at time $T$
is identical as for the matrix
\begin{equation}\label{DB-model}
H=H_0+\sqrt{T} H_1
\end{equation}
where $H_0=\diag(0,1,2,\ldots,N-1)$ and $H_1$ is an $N\times N$ matrix from the 
Gaussian Unitary Ensemble, with the second moment $\langle\tr H_1^2\rangle=N^2$.
The joint probability density of the eigenvalues $E=\diag(E_1,\ldots,E_N)$ of $H$ 
is given by
\begin{equation}\label{jpdfDB}
{P_H(E)=\frac{\exp[-\sum_{j=0}^{N-1}j^2/(2T)]}{N!\prod_{j=0}^{N-1}(\sqrt{2\pi T}\ j!)}\Delta_N\left(E\right)\Delta_N\left(e^{E/T}\right)\exp\left[-\frac{\tr E^2}{2T}\right]},
\end{equation}
where $\Delta_N(E)=\det[E_a^{b-1}]_{a,b=1,\ldots N}$ is the Vandermonde determinant.
The probability density~\eqref{jpdfDB} can be expressed as
a determinant of the kernel (see Eq.~(3.14) in \cite{KurtBrown})
\begin{equation}\label{kernel.DB}
\begin{split}
K_H^{(N)}(E_1,E_2)=&\sum_{j=0}^{N-1}e^{-(j-E_1)^2/(2T)}\int_{-\infty}^\infty\frac{dw}{2\pi T} e^{(iw-E_2)^2/(2T)}\prod_{l\neq j}\frac{iw-l}{j-l}\\
=&\sum_{j=0}^{N-1}\int_{-\infty}^\infty\frac{dw}{2\pi T} \frac{\sin[\pi(j-iw)]}{\pi(j-iw)}e^{(iw-E_2)^2/(2T)-(j-E_1)^2/(2T)}\frac{\Gamma[N-iw]\Gamma[iw+1]}{\Gamma[N-j]\Gamma[j+1]}.
\end{split}
\end{equation}
{We would like to mention that in~\cite{KurtBrown} the initial conditions were symmetrically chosen about the origin while in our case the origin is at the position of the lowest eigenvalue. Apart from this trivial shift the kernel~\eqref{kernel.DB} is the one in~\cite{KurtBrown}.}

One can study the limiting forms of the kernel in the local scale at the edges and in the bulk. In particular it was shown in \cite{KurtBrown} that the kernel takes the following form in the bulk 
\begin{equation}\label{kernel-trans-bulk.c}
\tilde{K}_{\rm cb}(\widehat{x},\widehat{y})= \frac{1}{\pi}\sum_{m=-\infty}^\infty \RE\left[\frac{\exp\left[-2\pi^2am(m-1)+i\pi [(\nu-\widehat y)+(2m-1)(\nu-\widehat x)]\right]}{2\pi m a+i(\widehat{x}-\widehat{y})}\right].
\end{equation}
The simplest way to derive this result from the kernel \eqref{kernel.DB} is to zoom in at the center of the spectrum where it is locally flat. {One can do this by setting $N=2n+1$ and choosing the base point close to $n$, i.e., $E_1=n+\nu+\widehat{x}$, $E_2=n+\nu+\widehat{y}$  and  $\nu\in]-0.5,0.5[$. Employing the expansion $w=n+\delta w$ and $j=n+m$ for $\delta w,m=\mathcal{O}(1)$ and the identity~\eqref{sine-identity}, one arrives after some manipulations at \eqref{kernel-trans-bulk.c} as was already done in \cite{KurtBrown}.
}

It has been surprising for us to discover that the kernel for Dyson's Brownian motion~\eqref{kernel-trans-bulk.c} is equivalent to the kernel for the product of Ginibre matrices \eqref{kernel-trans-bulk.ab} that we discussed in Section \ref{sec:trans.bulk}. The equivalence can be derived by applying the 
Poisson summation formula to \eqref{kernel-trans-bulk}. Writing
\begin{equation}
\begin{split}
F(h)=&\int_{-\infty}^\infty  ds\ {\rm erfi}\left[\pi\sqrt{\frac{a}{2}}+i\sqrt{\frac{1}{2a}}(s+\nu-\widehat{y})\right]\exp\left[\frac{1}{a}(\widehat{x}-\widehat{y})s-i h s\right]\\
=&\frac{2}{h+i(1/a)(\widehat{x}-\widehat{y})}\exp\left[-\frac{a}{2}\left(h+i\frac{1}{a}(\widehat{x}-\widehat{y})\right)^2+a\left(h+i\frac{1}{a}(\widehat{x}-\widehat{y})\right)\left(\pi+i\frac{1}{a}(\nu-\widehat{y})\right)\right],
\end{split}
\end{equation}
we have
\begin{equation}
\begin{split}
K_{\rm cb}(\widehat{x},\widehat{y})=&\frac{1}{2\pi a}\sum_{m=-\infty}^\infty \RE[F(2\pi m)]\\
=&\frac{1}{\pi a}\sum_{m=-\infty}^\infty \RE\left[\frac{\exp\left[-\frac{a}{2}\left(\pi (2m-1)+i\frac{1}{a}(\widehat{x}-\nu)\right)^2+\frac{a}{2}\left(\pi +i\frac{1}{a}(\nu-\widehat{y})\right)^2\right]}{2\pi m+i(1/a)(\widehat{x}-\widehat{y})}\right]\\
=&\frac{\exp\left[\frac{1}{2a}\left(\widehat{x}-\nu\right)^2-\frac{1}{2a}\left(\nu-\widehat{y}\right)^2\right]}{\pi}\sum_{m=-\infty}^\infty \RE\left[\frac{\exp\left[-2\pi^2am(m-1)+i\pi [(\nu-\widehat y)+(2m-1)(\nu-\widehat x)]\right]}{2\pi ma+i(\widehat{x}-\widehat{y})}\right]\\
=&\exp\left[\frac{1}{2a}\left(\widehat{x}-\nu\right)^2-\frac{1}{2a}\left(\nu-\widehat{y}\right)^2\right] \tilde{K}_{\rm cb}(\widehat{x},\widehat{y}).
\end{split}
\end{equation}
We see that up to an irrelevant factor $\exp\left[\left(\widehat{x}-\nu\right)^2/(2a)-\left(\nu-\widehat{y}\right)^2/(2a)\right]$, {see~\eqref{gauge}}, the two kernels are identical. The identification of time $T$ in the Dysonian Brownian Motion with the parameter $M/z_0$ is rather straightforward, namely $T=a=z_0/M$. The number of matrices $M$ in the product is proportional to time, if one interpretes the product as a transfer matrix, but we also see, that it is inversely proportional to time when one maps the kernel to that of Brownian Motion. It is a sort of duality. For increasing $M$ the picket fence statistics crystallises~\cite{ABK}. 
The duality manifests also as a map between real space modes in one picture and Fourier modes in the other one which is provided by the Poisson summation formula.

It is worth mentioning that a relation between the Dysonian Random Walk and a multiplicative stochastic process can be found also in \cite{IS} where the joint probability density for singular values of a product matrix is identical to~\eqref{jpdfDB}.

{As we have demonstrated above, the local statistics of the singular values 
of the random matrix product in the bulk is described by the same kernel for Dysonian Brownian Motion with the initial condition given by a picket fence. We can extend that statement even to the soft edge. For this purpose, we study the behaviour of the
kernel~\eqref{kernel.DB} for energy levels near the upper edge of the spectrum  
\begin{equation}\label{DB-soft-scale}
E_1=T(\chi+{\rm ln}(N))\ {\rm and}\ E_2=T(\zeta+{\rm ln}(N))\ {\rm with}\ \chi,\zeta=\mathcal{O}(1).
\end{equation}
We choose the summation index $j=N-1-\widehat{j}$ and the integration variable is $iw=N-1-\widehat{j}+i\widehat{t}$ with $\widehat{j},\widehat{t}=\mathcal{O}(1)$, {in~\eqref{kernel.DB}.} The ratio of the Gamma functions can be approximated like $\Gamma[N-\widehat{j}+i\widehat{t}]/\Gamma[N-\widehat{j}]\approx N^{it}$ which cancels with the logarithmic shift in the spectral variables, cf., Eq.~\eqref{DB-soft-scale}. The other Gamma function $\Gamma[1+\widehat{j}-i\widehat{t}]$ can be rewritten to $(-1)^{\widehat{j}}\pi /(\Gamma[i\widehat{t}-\widehat{j}]\sin[i\pi\widehat{t}])$ via Euler's reflection formula~\eqref{reflect} which also cancels the resulting sine function. The binomial series~\eqref{binomial.trick} leads to
\begin{equation}
\lim_{N\to\infty} T\frac{\exp[(\chi+{\rm ln}(N)-(N-1)/T)^2]}{\exp[(\zeta+{\rm ln}(N)-(N-1)/T)^2]}K_H^{(N)}(T(\chi+{\rm ln}(N)),T(\zeta+{\rm ln}(N)))=K_{\rm cs}\left(\chi-\frac{1}{2T},\zeta-\frac{1}{2T}\right)
\end{equation}
showing that the Dyson Brownian motion with $T=a=N/M$ gives the same limiting result 
at the soft edge as the multiplicative model discussed here. The critical behaviour at the soft edge shown above was not discussed in~\cite{KurtBrown}, so it is a new result.}

%%%%%%%%%%%%%%%%%%%%%%%%%%%%%%%%%%%%%%%%%%%%%%%%%%
\section{Numerical Simulations and Universality}\label{sec:universality}

It is tempting to conjecture that the local statistics is universal, 
 which means
that it holds not only for the product of independent Gaussian matrices, but for a wider class of multiplicative stochastic processes in matrix space.
In order to support this conjecture we have performed Monte Carlo simulations of six different kinds of products of random matrices. They are listed below.

\begin{figure}[t!]
 \centering
\includegraphics[width=1\textwidth]{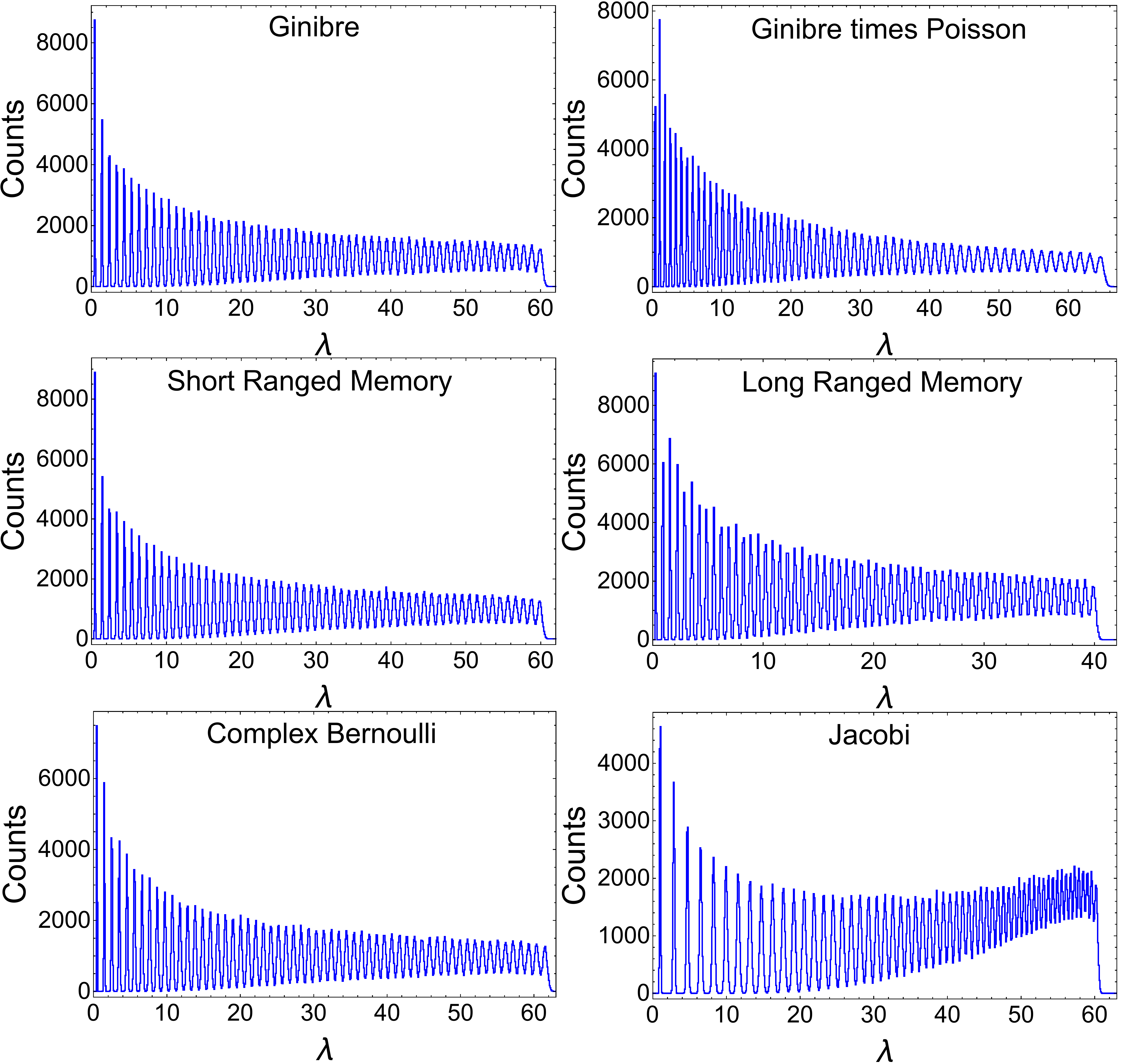}
 \caption{The total eigenvalue counts {({histogram values are} proportional  to the level density)} of the six ensembles. The bin size is equal to $0.1$ {and has been} chosen to be significantly smaller than the peaks' widths in order
to see local fluctuations of the level density. Each ensemble consists of $10^4$ configurations of a product of $M=500$ matrices of size $60\times 60$. Therefore, each histogram comprises $6\times 10^5$ eigenvalues. We have slightly rescaled the eigenvalues to {be able to apply} the same bin size.}
 \label{fig:numdensities}
\end{figure}

\begin{enumerate}
\item	The product of independent and identically distributed complex Ginibre matrices, see~\eqref{Ginibre-P}, has been the main object of interest for our analytical study. Thus it is most natural to consider it as our first ensemble to simulate, meaning we compute the squared singular values of the product matrix $X^{(M)}$ that is recursively defined by
			\begin{equation}\label{prod-num}
			X^{(j)}=X_j X^{(j-1)}\quad {\rm with}\quad X^{(1)}=X_1\ , \quad {\mbox{for}\ j=2,\ldots,M,}
			\end{equation}
			and each $X_j$ is drawn from the Gaussian distribution~\eqref{Ginibre-P}.
\item	As a second product matrix, we generate $X^{(M)}$ as in~\eqref{prod-num}, except that each $X_j$ is itself a product
			\begin{equation}
			X_j=\tilde{X}_jD_j,
			\end{equation}
			where $\tilde{X}_j$ are independent Ginibre matrices drawn from~\eqref{Ginibre-P} and $D_j$ are independent diagonal matrices whose diagonal matrix entries are independently and uniformly drawn from the interval $[0.5,1.5]$. Thence, $X^{(M)}$ is an alternating product of complex Ginibre matrices and real diagonal matrices, that would alone yield Poisson spectral statistics.

\begin{figure}[t!]
 \centering
\includegraphics[width=1\textwidth]{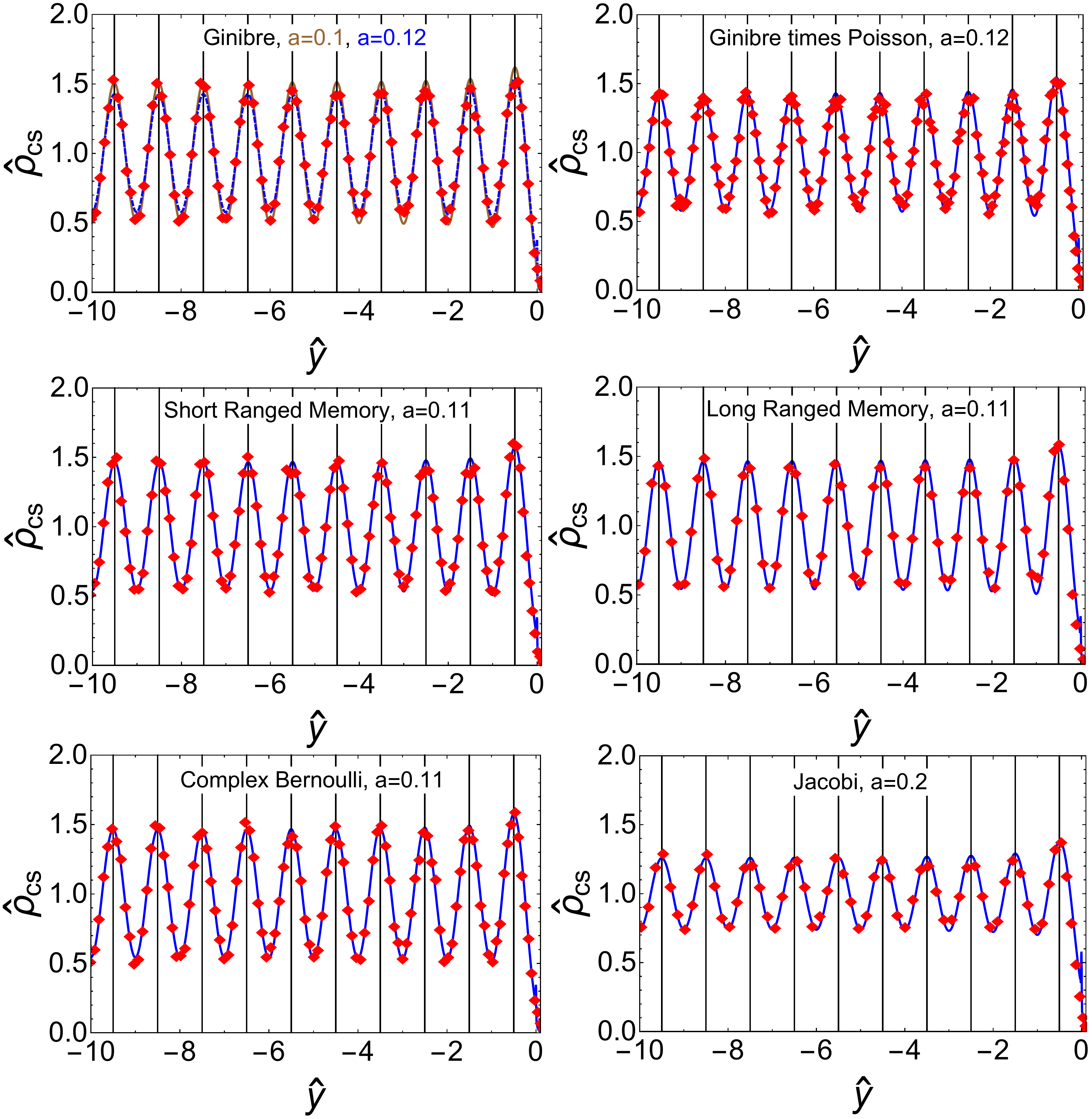}
 \caption{The unfolded microscopic level density of the six ensembles {1.-6.} of product matrices at the soft edge (symbols) compared to the analytical result~\eqref{crit-soft-unfold} in combination with~\eqref{kernel.critical.a} (smooth curves), with a fitted effective value $a$ that replaces the ratio $N/M$. Even for the Ginibre case (upper left plot), for which we have analytically studied the double scaling limits, we cannot rely completely on the relation $a=N/M$, which would be $a=60/500=0.12$ (blue dashed curve) in the present case. Due to the finite matrix size some deviations show {up} for smaller eigenvalues. Therefore, we have additionally plotted the result for $a=50/500=0.1$ (brown curve) which fits the tenth largest eigenvalue best. Certainly, similar deviations for the other ensembles can be explained by finite size effects, and should be always taken into account when comparing with empirical data. The matrix size {is} $N=60$, the number {of matrices multiplied is} $M=500$, and the ensemble size is $10^4$ for all six ensembles.}
 \label{fig:numsoft}
\end{figure}

\item	The {next} matrix product $X^{(M)}$, that {as a first example} involves correlations, is constructed via the recursive relation
			\begin{equation}
			X^{(j)}=(X_j+X_{j-1}) X^{(j-1)}\quad {\rm with}\quad {X^{(1)}=X_1\ , \quad \mbox{for}\ j=2,\ldots,M,}
			\end{equation}
			with independent $X_j$ drawn from~\eqref{Ginibre-P}. We call it short ranged memory model since the consecutive matrices in the product $X_j+X_{j-1}$ and $X_{j+1}+X_j$ are correlated through $X_j$, which contributes to both of them.
\item	A longer ranged memory model is the one with
			\begin{equation}
			X^{(j)}=\left(\sum_{l=1}^j2^{l-j}X_l\right) X^{(j-1)}\ ,\quad {\rm with}\quad X^{(1)}=X_1\ , \mbox{for}\ j=2,\ldots,M,
			\end{equation}
			and independently distributed $X_j$ drawn from~\eqref{Ginibre-P}. The factor $2^{l-j}$ exhibits an exponential decay of the correlation of the new matrix multiplied with respect to the past ones.
\item	To emphasise that also non-Gaussian ensembles share the same limiting statistics, we employed the recursion~\eqref{prod-num}, but now with independent complex Bernoulli matrices $X_j$, meaning each matrix entry of $X_j$ is independently and uniformely drawn from the set $\{0,\pm1\}+i\{0,\pm1\}$.
\item	Another ensemble which now destroys the independence of the matrix entries, but not the one between the matrices $X_j$,  is the Jacobi ensemble. Such a Jacobi ensemble can be generated by taking a sub-block, which is $X_j$, of a Haar-distributed unitary random matrix $U_j\in\U(L)$ with $L>N$. In the present case, we have chosen $L=2N$. We have anew exploited the recursion~\eqref{prod-num} and have drawn $M$ independent $U_j\in\U(2N)$ so that also the sub-blocks $X_j$, that are multiplied, are independent. {Recently, this product has been analytically studied in~\cite{A}.}
\end{enumerate}

The level density (times the  number of matrices generated) of the six ensembles is drawn in Fig.~\ref{fig:numdensities}. For all ensembles we have set $M=500$ and $N=60$, and generated $10^4$ product matrices. In some ensembles, we have rescaled the eigenvalues by a constant factor. In this way, we could choose the same bin size.

The peaks of each single eigenvalue can be nicely seen in each plot in Fig.~\ref{fig:numdensities}. This picture is very natural and shows that the macroscopic and mesoscopic level densities are not approached uniformly, when being in the critical ($a=\mathcal{O}(1)$) or in the sub-critical ($a\ll1$) regime. The oscillations are persistent. 

Additionally, it is evident that the hard edge about the origin always converges to picket fence statistics, regardless whether the matrices are Gaussian or non-Gaussian, and uncorrelated or correlated. The distributions of the individual eigenvalues only start to overlap significantly when departing from the origin and have the strongest overlap at the soft edge.

It is the soft edge result~\eqref{crit-soft-unfold} that we fit to the numerics, with an effective $a$ that has not necessarily to be $N/M$ for the non-Ginibre ensembles. This is particularly seen for the product of complex Jacobi matrices, where $a\approx0.2$ while $M$ and $N$ are still the same as the Ginibre case. Albeit we have chosen the soft edge and not the bulk, the fitting with the soft edge result highlights also the agreement with the bulk statistics~\eqref{kernel-trans-bulk}. The transition from the soft edge to the bulk statistics is rapid, as we know from the transition between the Airy-kernel and the sine-kernel. Already after three or four eigenvalues the statistical error of the empirical data is larger than the actual deviation between the two statistics.

When fitting empirical data with our analytical result {in Fig. \ref{fig:numsoft}}, one has to be aware of two things. Firstly, the finite size effects can be significant and visible. This can be easily observed in the Ginibre case (upper left plot in Fig.~\ref{fig:numsoft} where the parameter $a$ is for the largest eigenvalue $a=N/M=60/500=0.12$ while for the tenth largest eigenvalue we have $a=j/M=50/500=0.1$. These deviations have to be taken into account, especially when there are no analytical formulas at hand for the macroscopic or mesoscopic level density.

The next problem to solve is to fix the position of the largest eigenvalue. While the distributions of the other eigenvalues can be relatively easily fixed by enforcing that the mean level spacing of each pair of consecutive eigenvalues is equal to one, we have no straightforward information about the position of the largest eigenvalue. We have solved this problem by fixing the tail, by rescaling the distribution of the largest eigenvalue with a fixed constant. This very crude method works rather well when considering the results shown in Fig.~{\ref{fig:numsoft}.}

%%%%%%%%%%%%%%%%%%%%%%%%%%%%%%%%%%%%
\section{Summary}\label{sec:conclusio}

In the present work we have delivered  the details of the derivation for the results for the double scaling limits of a product of Ginibre matrices, presented by us in the letter~\cite{ABK}. Furthermore, we gained more insights in the mechanisms behind the transition of the local spectral statistics from picket fence (equidistant eigenvalues) to GUE statistics. One of these insights is the exact equivalence of the kernel with those that can be obtained from Dyson's Brownian motion, with the picket fence statistics as its initial condition, which we have proven in the present work for the bulk as well as the soft edge statistics. The main difference of the additive process of Dyson's Brownian motion and of its multiplicative counterpart considered here, is that the transition of the statistics is not uniform for the whole spectrum. In general different parts of the spectrum of such a product matrix correspond to different times in the Dyson Brownian motion. This time parameter is equal to $a=j/M$, where $j$ stands for the $j$-th smallest eigenvalue, or in general equal to the squared width to spacing {ratio} $a={\rm WSR}_j^2$, cf., Eq.~\eqref{WSRdef}.

This kind of universality between additive and multiplicative stochastic processes on matrix spaces {was} 
substantiated with Monte-Carlo simulations of product matrices, that also comprise correlations between the matrices multiplied, and non-Gaussian ensembles. We have fitted the analytical results at the soft edge to the empirical data, {finding very good agreement}. In doing so, we have exploited another new insight which concerns an emerging mesoscopic spectral scale {interpolating} between the bulk and the soft edge. Albeit the microscopic statistics in this very narrow part close to the largest eigenvalue is the same {as} the bulk statistics, its averaged level density does not follow the one from the bulk. It is exactly the mesoscopic level density that is needed to properly unfold the spectrum at the soft edge since it always exhibits a square root behaviour. This square root {edge} is not seen any more, regardless whether and how the matrix dimension $N$ and the number $M$ of matrices multiplied are related. An open question is whether the mesoscopic wide correlators will be also very different from those of the macroscopic scale.

{Several generalisations of  products of Ginibre matrices have been studied such as products of non-Gaussian ensembles~\cite{kieburg,KKS,A,akemannstrahov} as well as of products of rectangular matrices~\cite{AIK,kuijlaars,kuijlaars2}. Regarding the first, there was recently a study~\cite{A} on products of Jacobi ensembles (truncated unitary matrices) where the authors found similar effects as we have seen, see also Sec.~\ref{sec:universality}. We also expect that this carries over to a product over rectangular matrices where in the end  only the average of the rectangularity $\nu$ (difference of the two matrix sizes) of the matrices multiplied may enter the game. We have not considered those two generalisations in the present work to keep the technical level of our results as transparent as possible.}

Another direction in which one can try to extend the ideas presented in our work is to {consider} products of real and quaternionic Ginibre matrices. Do those ensembles yield local spectral statistics that still follow those of Dyson's Brownian motion but now for the Gaussian orthogonal and symplectic ensemble, respectively? It is also interesting to study {complex} eigenvalues (instead of singular values) of the product of Ginibre matrices using the tools developed in {\cite{ab}, where we refer to \cite{LW} for first results}. Due to its two-dimensional nature one should expect a different but related behaviour due to the equivalence of the eigenvalue and singular value statistics for bi-unitarily invariant random matrix ensembles that was proven in~\cite{KK16}. Products of Ginibre ensembles surely satisfy the requirements that are needed for this kind of equivalence.

\section{Acknowledgments}

This work was supported by the German Science
Foundation (DFG), through grant CRC1283 ``Taming uncertainty and profiting
from randomness and low regularity in analysis, stochastics and their
applications'' (GA and MK). The 
Faculty of Physics and Applied Computer Science at 
AGH University of Science and Technology is thanked for hospitality (GA and MK) as well as the 
School of Mathematics and Statistics of the University of Melbourne (GA).  

%%%%%%%%%%%%%%%%%%%%%%%%%%%%%%%%%%%%%%%%%%%%%%%%%%%%%%%
\appendix
\section{Derivation of the Kernel Representation}
\label{app:main}

In this appendix, we derive the kernel 
given in \eqref{eq:main1a} and \eqref{eq:main1b}.
As a starting point we use the representation of the kernel in terms of Meijer-G functions given in \cite{AIK}, 
\begin{equation}
\label{G-kernel}
K_Y(x,y)=
\sum_{j=1}^{N} \MeijerG{1}{0}{1}{M+1}{j}{0,\,\ldots\,,0}{x} 
\MeijerG{M}{1}{1}{M+1}{-j+1}{0,\,\ldots\,,0}{y}\ .
\end{equation}
While for the general definition of  Meijer-G functions we refer to \cite{NIST},  in this particular case the two functions in the last equation are given by the following complex contour integral representations:
\begin{equation}
K_Y(x,y)= \sum_{j=1}^N \int_{\gamma_s^\prime} \frac{ds}{2\pi i} x^s 
\frac{\Gamma(-s)}{\Gamma^M(1+s) \Gamma(j-s)} \int_{\gamma_t}
\frac{dt}{2\pi i} y^t \frac{\Gamma(j+t)\Gamma^M(-t)}{\Gamma(1+t)} \ .
\end{equation}
The contour $\gamma_s^\prime$ in the first integral encloses all poles of the integrand, given by the non-negative integers, in clockwise direction.
The second contour $\gamma_t$ is a straight line parallel to the imaginary axis, $c+i\mathbb{R}$, with $-j<c<0$ in between the poles of the integrand for the $j$-th term. In order to make the contour $j$-independent we have choosen $-1<c<0$.

It is convenient to first evaluate the sum over $j$. To that end, we regroup the terms as follows
\begin{equation}
\label{G-kernel2}
K_Y(x,y)= \int_{\gamma_s^\prime} \frac{ds}{2\pi i} x^s 
\frac{\Gamma(-s)}{\Gamma^M(1+s)} \int_{\gamma_t}
\frac{dt}{2\pi i} y^t \frac{\Gamma^M(-t)}{\Gamma(1+t)} \sum_{j=1}^N  \frac{\Gamma(j+t)}{\Gamma(j-s)}\ ,
\end{equation}
and apply the following telescopic property from \cite{kuijlaars2}, valid for general  integers $n_+\geq n_-$, 
\begin{equation}
\sum_{n=n_-}^{n_+}\frac{\Gamma(z+n)}{\Gamma(w+n)} = 
\frac{1}{z-w+1} \left(\frac{\Gamma(z+1+n_+)}{\Gamma(w+n_+)} - 
\frac{\Gamma(z+n_-)}{\Gamma(w-1+n_-)} \right) .
\label{eq:telescope}
\end{equation} 
It enables us to carry out the sum over $j$ in \eqref{G-kernel2}
\begin{equation}
K_Y(x,y)= \int_{\gamma_t}
\frac{dt}{2\pi i} y^t \frac{\Gamma^M(-t)}{\Gamma(1+t)} 
\int_{\gamma_s^\prime} \frac{ds}{2\pi i} x^s 
\frac{\Gamma(-s)}{\Gamma^M(1+s)} 
\frac{1}{t+s+1}\left\{\frac{\Gamma(N+t+1)}{\Gamma(N-s)} - \frac{\Gamma(1+t)}{\Gamma(-s)}\right\} \ .
\end{equation}
The second term in the curly brackets cancels all residua coming from 
$\Gamma(-s)$. Hence, the corresponding integral over $s$ vanishes if 
the pole at $t+s+1$ is not included in 
the contour $\gamma_s^\prime$. Under this condition, the remaining part can be written as
\begin{equation}\label{interm.kernel}
K_Y(x,y)= \int_{\gamma_t}
\frac{dt}{2\pi i} y^t \frac{\Gamma^M(-t)}{\Gamma(1+t)} 
\int_{\gamma_s^\prime} \frac{ds}{2\pi i} x^s 
\frac{\Gamma(-s)}{\Gamma^M(1+s)} 
\frac{1}{t+s+1} \frac{\Gamma(N+t+1)}{\Gamma(N-s)}  \ .
\end{equation}
This is the starting point for the two integral representations in 
\eqref{eq:main1a} and \eqref{eq:main1b}.

The first representation is obtained as follows.
The integral over $s$ picks up the contributions from the residua ${\rm Res}\, \Gamma(z)|_{z=-j} = (-1)^j/\Gamma(1+j)$ of the poles at $s=j$ for $j=0,1,\ldots, N-1$, yielding
\begin{equation}\label{interm.kernel2}
K_Y(x,y)= \int_{\gamma_t}
\frac{dt}{2\pi i} y^t \frac{\Gamma^M(-t)}{\Gamma(1+t)} \sum_{j=0}^{N-1}
\frac{(-1)^j x^j}{\Gamma(1+j)^{M+1}} \frac{1}{t+j+1} \frac{\Gamma(N+t+1)}{\Gamma(N-j)}\ .
\end{equation}
We employ Euler's reflection formula~\cite{NIST},
\begin{equation}\label{reflect}
\Gamma(z)\Gamma(1-z) = \frac{\pi }{\sin(\pi z)}\ ,\quad z\notin \mathbb{Z\ ,}
\end{equation}
to replace $1/\Gamma(1+t)$ by $-\Gamma(-t) \sin(\pi t)/\pi$. Finally, we change the
variable $-t \leftrightarrow t+1$ and we arrive at
\begin{equation}
K_Y(x,y)= \frac{1}{y}\sum_{j=0}^{N-1} x^j \int_{\gamma_t} \frac{dt}{2\pi i} 
y^{-t}\frac{\sin(\pi(j-t))}{\pi(j-t)} \left( \frac{\Gamma(1+t)}{\Gamma(1+j)}\right)^{M+1}\frac{\Gamma(N-t)}{\Gamma(N-j)}\ ,
\end{equation}
after making the contours $j$-independent by choosing $\gamma_t$ again.
The last equation can be cast into \eqref{eq:main1a}, 
\begin{equation}\label{Gexp}
\begin{split}
K_Y(x,y)= \frac{1}{y}\sum_{j=0}^{N-1} \int_{\gamma_t} \frac{dt}{2\pi i} \frac{\sin (\pi (j-t))}{\pi (j-t)}\ e^{i\pi {\rm sign}[\Im(t)]t} \exp\left[-\mathcal{S}(j;x)+\mathcal{S}(t;y)\right],
\end{split}
\end{equation}
when defining the action as 
\begin{equation}\label{Lagrangian}
\mathcal{S}(z;\alpha)=-i\pi \, {\rm sign}[\Im(z)]z-{\rm ln}[\alpha]z+(M+1){\rm ln} [\Gamma(1+z)]+ {\rm ln}\left[\Gamma(N-z)\right].
\end{equation}
This representation~\eqref{Gexp} is a good starting point when the local statistics show either picket fence statistics or are in the interpolating regime.

The second representation \eqref{eq:main1b}
%Another helpful equivalent representation to Eq.~\eqref{Gexp} 
can be obtained from~\eqref{interm.kernel}, by appling the reflection formula~\eqref{reflect} to the two Gamma functions $\Gamma(1+t)$ and $\Gamma(-s)$. After substituting $t+1 \to -t$ and $s\to -s$ we arrive at
\begin{equation}\label{Gexp.new}
\begin{split}
K_Y(x,y)= \frac{1}{y}\int_{\gamma_t} \frac{dt}{2\pi i} \int_{\gamma_s}\frac{d s}{2\pi i} \frac{1}{s-t}\frac{\sin(\pi t)\ e^{i\pi {\rm sign}[\Im(t)]t}}{\sin(\pi s)\ e^{i\pi {\rm sign}[\Im(s)]s}}\exp\left[-\mathcal{S}(s;x)+\mathcal{S}(t;y)\right].
\end{split}
\end{equation}
Notice that the contour $\gamma_s$ now runs in counter-clockwise direction and has been contracted, to only enclose the interval $[0,N-1]$. This is because the remaining poles  at larger positive integers of the sine-function in the denominator are cancelled by $\Gamma(N-z)$ from the action. 

A similar representation has been derived in~\cite{LWW}. It will become useful in the case where the local statistics follow the sine- or Airy-kernel. But we employ it also to obtain the macroscopic level density that is required to properly unfold the spectrum in Section \ref{sec:unfolding}. 

%%%%%%%%%%%%%%%%%%%%%%%%%%%%
\section{Saddle Point Analysis}
\label{app:sp}

The Digamma function $\psi(z) = \Gamma'(z)/\Gamma(z)$ has the following series representation \cite[5.76]{NIST} away from its poles $z=0,-1,-2,\ldots$
\begin{equation}
\label{digamma-series}
\psi(z)= -\gamma+\sum_{l=0}^\infty \left( \frac{1}{l+1}-\frac{1}{l+z}\right).
\end{equation}
Here, $\gamma$ is the Euler-Mascheroni constant. 
The solutions $z_{\rm s}$ of \eqref{saddlepointeqbulk} obviously come in complex conjugate pairs,  due to $\psi(z)^*=\psi(z^*)$, unless $z_{\rm s}\in\mathbb{R}$.
We will now show that we have a unique solution in the upper half plane of the complex plane, and thus also in the lower half plane. 
\begin{figure}[t!]
 \centering
\includegraphics[width=0.8\textwidth]{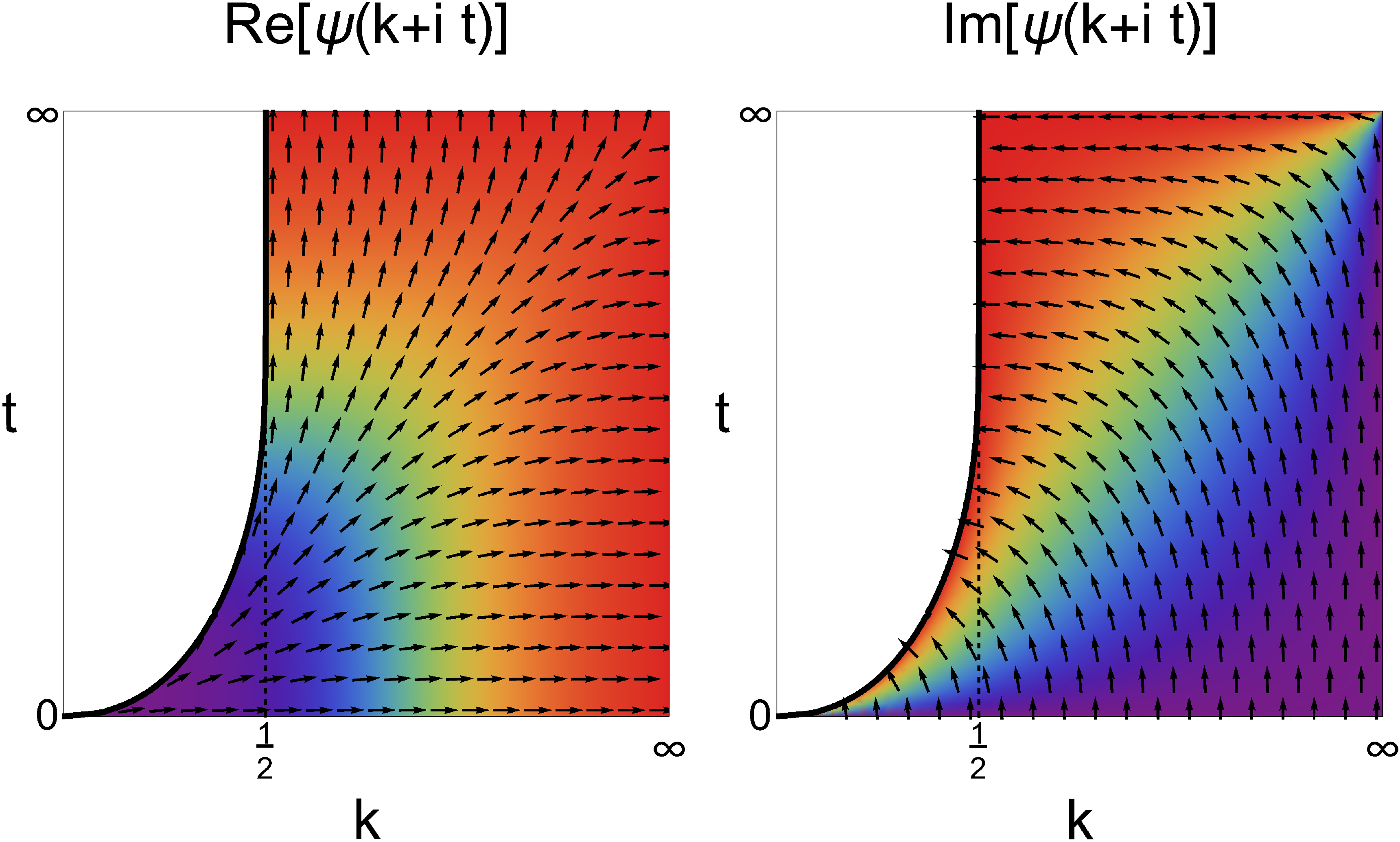}
 \caption{The real (left plot) and imaginary (right plot) part of the Digamma function~\eqref{im-Digamma} in the upper right quadrant of the complex plane. Note that we have mapped the whole quarter plane to a square by the monotonicity preserving map $(k,t)\mapsto(k/[1+k],t/[1+t])$. The black curve is defined by $\IM[\psi(k+i t)]=\pi/2$ and the dashed black line at $k=1/2$ indicates when the $\IM[\psi(k+i t)]<0$ is satisfied for all positive $t>0$. The height of the real and imaginary part is colour coded in a density plot where purple and red are either $-\infty$ and $+\infty$ for $\RE[\psi(k+i t)]$ or $0$ and $\pi/2$ for $\IM[\psi(k+i t)]$. The vectors highlight the local steepest ascend of the function, in particular they show our claim that $\IM[\psi(k+i t)]$ is increasing in $t$ and $\RE[\psi(k+i t)]$ is increasing in $k$ in this regime.}
 \label{fig:Digamma}
\end{figure}

Let us split $z=k+it$ into its real and imaginary part. First, we  fix the real part $\RE(z)=k>0$ of a complex solution and assume from now on that $\IM(z)=t\geq0$.
From \eqref{digamma-series}, it follows for the imaginary part of the Digamma function
\begin{equation}\label{im-Digamma}
\IM[\psi(k+i t)]=\sum_{l=0}^\infty\frac{t}{(k+l)^2+t^2},\quad \ k,t\in\mathbb{R}\ .
\end{equation}
It is a strictly increasing function in $t>0$ as long as $0\leq\IM[\psi(k+i t)]\leq\pi/2$, see Figure~\ref{fig:Digamma} (right plot). For $k\geq1/2$ the restriction $\IM[\psi(k+i t)]\leq\pi/2$ holds for all $t\in\mathbb{R}_+$. Unfortunately, we do not have any analytical proof for the monotonicity of the imaginary and real part of the Digamma function in this regime but only a strong numerical evidence. The condition $\IM[\psi(k+i t)]\leq\pi/2$ is important because the  imaginary part of~\eqref{saddlepointeqbulk} at the saddle point $z_{\rm s}=k+it$,
\begin{equation}
\label{Im-SP}
0=\IM[\partial_{z_{\rm s}}\mathcal{S}(z_{\rm s};y)]=(M+1)\sum_{l=0}^\infty\frac{t}{(1+k+l)^2+t^2}+ \sum_{l=0}^\infty\frac{t}{(N-k+l)^2+t^2} - \pi\ {\rm for}\ 0<k<N,
\end{equation}
can only be satisfied in this regime. Indeed both sums, representing the imaginary part of the two Digamma functions involved, are positive for $t>0$. Thence, there is a unique solution $t_0(k)>0$ for each fixed $k\in]0,N[$ as both sums are strictly increasing in $t>0$. Let us insert this function $t_0(k)$ into the real part of the saddle point equation~\eqref{saddlepointeqbulk} 
\begin{equation}
\label{Re-SP}
0=\RE[\partial_{z_{\rm s}}\mathcal{S}(z_{\rm s};y)]=-\ln(y)+ (M+1)
\sum_{l=0}^\infty\frac{(l+k+1)(k+2)+t^2}{(l+k+1)^2+t^2}-
\sum_{l=0}^\infty\frac{(l+N-k)(N-k+1)+t^2}{(l+N-k)^2+t^2}
\end{equation}
using 
\begin{equation}\label{re-Digamma}
\RE[\psi(k+i t)]=-\gamma + \sum_{l=0}^\infty\left[\frac{1}{l+1}-\frac{l+k}{(l+k)^2+t^2}\right]=-\gamma + \sum_{l=0}^\infty\frac{(l+k)(k-1)+t^2}{(l+1)[(l+k)^2+t^2]} .
\end{equation}
The latter is also strictly increasing for a strictly increasing function in $k>1/2$ for $t\geq0$, see left plot in Figure~\ref{fig:Digamma}, and thus we obtain also a unique solution for $k_0$ along $t_0(k)$.

Equation~\eqref{im-Digamma} is also ideal to get an idea of what order the imaginary part of $z_{\rm s}$ is in $N$ and $M$. 
Since the summand in Eq.~\eqref{im-Digamma} is a strictly decreasing function in the summation index $l$ for $t>0$, we can give the upper and lower limits for the sum
\begin{equation}
\int_0^\infty du \frac{t}{(k+u)^2+t^2}\leq \sum_{l=0}^\infty\frac{t}{(k+l)^2+t^2}
\leq \int_{-1}^\infty du \frac{t}{(k+u)^2+t^2}\ .
\end{equation}
After evaluating these integrals and using $\arctan(1/x)=\mbox{arccot}(x)=\frac{\pi}{2}-\arctan(x)$ for $x>0$, we have \eqref{im-Digamma}
\begin{equation}
\label{int-bound}
{\arctan}\left(\frac{t}{k}\right) \leq \IM[\psi(k+i t)] \leq  {\arctan}\left(\frac{t}{k-1}\right),
\end{equation}
with $k>1$ for the upper bound and $k>0$ for the lower one. 
Inserting the lower and the upper bound of \eqref{int-bound} into \eqref{Im-SP}, we obtain
\begin{equation}
\label{lower}
(M+1)\arctan\left( \frac{t}{1+k}\right)+ \arctan\left(\frac{t}{N-k}\right) -\pi\leq 0\ ,
\end{equation}
and
\begin{equation}
\label{upper}
(M+1)\arctan\left( \frac{t}{k}\right)+ \arctan\left(\frac{t}{N-k-1}\right) -\pi\geq 0\ .
\end{equation}
respectively.
Simple manipulations of \eqref{lower} and \eqref{upper} lead to
\begin{equation}
\begin{split}\label{imaginary-bounds}
\IM[z_{\rm s}]\leq &\ (1+\RE[z_{\rm s}])\tan\left[\frac{1}{M+1}\left(\pi-{\rm arctan}\left(\frac{\IM(z_{\rm s})}{N-\RE[z_{\rm s}]}\right)\right)\right]\,\leq (1+\RE(z_{\rm s}))\tan\left[\frac{\pi}{M+1}\right],\\
\IM[z_{\rm s}]\geq &\ \RE[z_{\rm s}]\tan\left[\frac{1}{M+1}\left(\pi-{\rm arctan}\left(\frac{\IM(z_{\rm s})}{N-1-\RE[z_{\rm s}]}\right)\right)\right]\quad\geq \RE[z_{\rm s}] \tan\left[\frac{\pi}{2(M+1)}\right],
\end{split}
\end{equation}
where we have reinserted $k=\RE[z_{\rm s}]$ and $t=\IM(z_{\rm s})\geq0$.
For the inequalities on the right hand sides we have used the monotonicity of $\arctan$, and the fact that it is bounded by $0$ and $\pi/2$ for positive arguments.

\section{Resolvent -- {Saddle Point Relation for $\RE[z_{\rm s}]\gg M$}} \label{app:resolvent}

{The saddle point $z_{\rm s}$  can be related to the limiting resolvent $G_Y$ using \eqref{level.saddle}.} It is defined as an integral of the limiting normalised density $\rho_Y$ over its support $\sigma$
\begin{equation}
\label{GYdef}
G_Y({w})=\lim_{N\to\infty}\int_{w \notin \sigma} \frac{du \ \rho_Y(u)}{{w}-u} \overset{|w|\gg 1}{\approx}\frac{1}{w}\ .
\end{equation}
The large argument asymptotic follows from the normalisation of the density. 
The limiting density is recovered via
\begin{equation} \label{rhoG}
\rho_Y({y})=-\frac{1}{2\pi i}\lim_{\epsilon\searrow 0} 
[G_Y({y}+i\epsilon)-G_Y({y}-i\epsilon)]=\frac{1}{\pi}\lim_{\epsilon\searrow 0} \IM G_Y({y}-i\epsilon)\ .
\end{equation}
Comparing with the  relation~\eqref{level.saddle}, this suggests to identify 
\begin{equation}
\label{zs-resolvent}
z_{\rm s}(y)\overset{M,N\gg1}{\approx}N\frac{y}{N^M} \lim_{\epsilon\searrow 0} 
G_Y\left(\frac{y}{N^M}-i\epsilon\right) 
\end{equation}
with the resolvent for the matrix $Y/N^M$,
where we have inserted the scaling implied by \eqref{zs.def}. The large argument behaviour of $z_{\rm s}(y)$ found above is consistent with that of the resolvent in \eqref{GYdef} and fixes all constants. 

The scaling considered here is for $N\gg M$, {as follows from \ref{sec:densityggM}}. Let us therefore make contact with previous results for $\rho_Y$ obtained in the limit $N\to\infty$ with fixed $M$, as here the limiting density and resolvent are known~{\cite{BLS,TN}}.
We denote by $G^{(M)}$ 
the limiting Green function 
of the product matrix $Y/N^M$ when we take $N\to\infty$ for a fixed $M$, i.e.,
\begin{equation}\label{green-M}
G^{(M)}(\widehat{z})=\int_0^{(M+1)^{M+1}/M^M}\frac{\rho^{(M)}(\zeta)d\zeta}{\widehat{z}-\zeta}.
\end{equation}
The limiting density fulfils the equation \cite{BLS}.
\begin{equation}
\widehat{z}^M \left(G^{(M)}\left(\widehat{z}\right)\right)^{M+1} - 
\widehat{z}\, G^{(M)}\left(\widehat{z}\right) + 1 =0
\end{equation}
The corresponding limiting level density $\rho^{(M)}(\zeta)$
was found analytically in \cite{TN}, {which} 
is given in Eq. 
\eqref{level.finiteM}.

%%%%%%%%%%%%%%%%%%%%%%%%%%%%%%%%%%%%%%%%%%%%%%%%%%%
\section{Distance Between $\mathbf{z_0}$ and $\mathbf{z_{\rm s}}$ {when Re$[z_s]\leq \mathcal{O}(M)$}} \label{app:z0zs}

In a similar fashion as in Eqs. \eqref{difference} and \eqref{difference.b}, 
one can also estimate the real part of the difference of the action at $z_0$ and at the original saddle point $z_{\rm s}$ in the complex plane, i.e.,
\begin{equation}\label{difference.c}
\begin{split}
\widetilde\Delta=&\RE[\mathcal{S}(z_{\rm s};y)-\mathcal{S}(z_{0};y)]=\RE\biggl[
(\Delta z)^2\int_0^1d\lambda(1-\lambda)\left[(M+1)\psi'(1+z_0+\Delta z\lambda)+\psi'(N-z_0-\Delta z\lambda)\right]\biggl],
\end{split}
\end{equation}
with $\Delta z=z_{\rm s}-z_0$. The value of the quantity $\widetilde\Delta$ will tell us whether the approximation of the original saddle point by $z_0$ is legitimate.

In order to get a feeling whether $\Delta z$ is large or small, we start with the initial saddle point equation \eqref{saddlepointeqbulk} and show that $\Delta z$ is maximally of order one. Making use of~\eqref{Imzs-order}, meaning $\IM[z_{\rm s}]=\mathcal{O}(\RE[z_{\rm s}]/M)$, we assume that we can expand the saddle point equation~\eqref{saddlepointeqbulk} in $\IM[z_{\rm s}]\leq \mathcal{O}(1)$, and then take the real part. The first term in the Taylor expansion vanishes since it is imaginary, so that we arrive at
\begin{equation}\label{saddle-approx}
-{\rm ln}(y)+(M+1)\psi(1+\RE[z_{\rm s}])-\psi(N-\RE[z_{\rm s}])+\mathcal{O}\left(\left[(M+1)\psi''(1+\RE[z_{\rm s}])-\psi''(N-\RE[z_{\rm s}])\right]\left[\frac{\RE[z_{\rm s}]}{M}\right]^2\right)= 0.
\end{equation}
Our assumption has been that we stay far away from the hard edge and soft edge, i.e., $1+\RE[z_{\rm s}]\gg1$ and $N-1-\RE[z_{\rm s}]\gg 1$, respectively, see~\eqref{bulk-cond}. Thus, we are allowed to approximate the Digamma functions and its derivatives by their leading terms, see~\eqref{asymptotic.Digamma}. Then, the correction of the saddle point equation in~\eqref{saddle-approx} is of order
\begin{equation}
\mathcal{O}\left(\Big[(M+1)\psi''(1+\RE[z_{\rm s}])-\psi''(N-\RE[z_{\rm s}])\Big]\left[\frac{\RE[z_{\rm s}]}{M}\right]^2\right)
=\mathcal{O}\left(
\mbox{max}\left\{\frac{1}{M}, \frac{\RE[z_{\rm s}]^2
}{M^2(N-\RE[z_{\rm s}])^2}\right\}\right)
\ll1. 
\end{equation}
Here,  we have exploited $\RE[z_{\rm s}]\leq\mathcal{O}(M)$ in the current situation, and that $M$ is large.
 
Let us come back to the question how far away we are with $z_0$ from the true saddle point $z_{\rm s}$. The solution of \eqref{saddle-real} is uniquely given by $z_0$, without any correction term. Expanding the first term in \eqref{saddle-approx} in the difference $\RE[\Delta z]=\RE[z_{\rm s}]-z_0$, which is small compared to $z_0$, we need to enforce that the leading correction in $\RE[\Delta z]$ has to cancel the second term shown in~\eqref{saddle-approx}. Consequently,  we get the following scaling relation from setting both orders to be equal,
\begin{equation}
\left[(M+1)\psi'(1+z_{\rm 0})+\psi'(N-z_{\rm 0})\right]\RE[\Delta z]\approx\left[\frac{M}{z_{\rm 0}}+\frac{1}{N-z_{\rm 0}}\right]\RE[\Delta z]=\mathcal{O}\left(\mbox{max}\left\{\frac{1}{M},\frac{\RE[z_{\rm s}]^2}{M^2(N-\RE[z_{\rm s}])^2}\right\}\right).
\end{equation}
Equivalently, it holds  
\begin{equation}
|\RE[\Delta z]|=\mathcal{O}
\left(\mbox{max}\left\{\frac{z_0}{M^2},\frac{z_0\RE[z_{\rm s}]^2}{M^3(N-\RE[z_{\rm s}])^2}\right\}\right)\ll1, 
\end{equation}
since $M$ is large, $\RE[z_{\rm s}]\leq \mathcal{O}(M)$, $z_0/M\leq \mathcal{O}(1)$,  and we are in the bulk \eqref{bulk-cond}.
{This implies  
$|\Delta z|=|z_{\rm s}-z_0|=\mathcal{O}(|\IM[z_{\rm s}]|)$, 
because the real part $\RE[z_{\rm s}]$ always dominates the imaginary part $\IM[z_{\rm s}]$, cf. \eqref{Imzs-order}.} Furthermore, the imaginary part $\IM[z_{\rm s}]=\mathcal{O}(\RE[z_{\rm s}]/M)$ is always bigger than $|\RE[\Delta z]|$, and thus it determines the order of $\Delta z$.

Summarising, we have not only found that the true saddle point $z_{\rm s}$ is close to $z_0$, but also that the real part $\RE[z_{\rm s}]$ converges to $z_0$ when $M,N\to\infty$. 
  
Next, we evaluate the difference $\widetilde\Delta$ in~\eqref{difference.c}. This can be done by expanding the original difference, \eqref{difference} with the replacement $j\to z_{\rm s}$, and inserting $z_{\rm s}=\Delta z+z_0$ to expand in $\Delta z$. The logarithms just give an expansion in the Digamma function and its derivatives, and after the cancellation in the first order  we obtain
\begin{equation}
\label{delta-exp}
\widetilde\Delta\approx (M+1)\frac12 (\Delta z)^2 \psi'(1+z_0)+\frac12 (\Delta z)^2 \psi'(N-z_0) \approx \frac12 (\Delta z)^2\left( \frac{M}{z_0}+ \frac{1}{N-z_0}\right). 
\end{equation}
Due to the condition \eqref{bulk-cond2}, we can  exploit the asymptotic formulas~\eqref{asymptotic.Digamma} for the derivative of the Digamma function in the second step.
Clearly, the first term in the bracket on the right hand side is larger than or equal to the order $ \mathcal{O}(1)$ and the second term is much less than one so we obtain
\begin{equation}
|\widetilde{\Delta}|=\mathcal{O}(z_0/M)\leq\mathcal{O}(1)
\end{equation}
agreeing with the order of $|\IM[z_{\rm s}]|$.
Hence, the expansion about $z_0$ for the $t$-integral as well as for the summation index $j$ in~\eqref{eq:main1a} about the point $z_0$ instead of $z_{\rm s}$ is justified because we cover all contributions from the saddle point.

%%%%%%%%%%%%%%%%%%%%%%%%%%%%%%%

\end{document}